\documentclass[12pt,preprint]{aastex}

\usepackage{subfigure}
\usepackage{enumerate}
\usepackage{color}
\begin{document}

\shorttitle{Synthesis of a simulated type~{\sc ii} spicule}
\shortauthors{Mart\'inez-Sykora \& De Pontieu \& Pereira \& Leenaarts \& Hansteen}

\title{A Detailed Comparison Between The Observed and Synthesized Properties of a Simulated Type~{\sc ii} Spicule}

\author{Juan Mart\'inez-Sykora$^{1,2,3}$}
\email{j.m.sykora@astro.uio.no}
\author{Bart De Pontieu$^{1}$} 
\author{Jorrit Leenaarts$^{2}$}
\author{Tiago M. D. Pereira$^{1,4}$}
\author{Mats Carlsson$^{2}$} 
\author{Viggo Hansteen$^{2}$} 
\author{Julie V. Stern$^{1}$} 
\author{Hui Tian$^{6}$}
\author{Scott W. McIntosh$^{5}$}  
\author{Luc Rouppe van der Voort$^{2}$} 
\affil{$^1$ Lockheed Martin Solar and Astrophysics Laboratory, Palo Alto, CA 94304, USA}
\affil{$^2$ Institute of Theoretical Astrophysics, University of Oslo, P.O. Box 1029 Blindern, N-0315 Oslo, Norway}
\affil{$^3$ Bay Area Environmental Research Institute, Sonoma, CA 95476, USA}
\affil{$^4$ NASA Ames Research Center, Moffett Field, CA 94035, USA}
\affil{$^5$ High Altitude Observatory, National Center for Atmospheric Research, Boulder, CO 80307, USA}
\affil{$^6$ Harvard-Smithsonian Center for Astrophysics, Cambridge, MA 02138, USA}

\newcommand{\myemail}{juanms@astro.uio.no}
\newcommand{\viscous}{\underline{\underline{\tau}}}
\newcommand{\resistive}{\underline{\underline{\eta}}}
\newcommand{\komment}[1]{\texttt{#1}}
\newcommand{\jmsk}[1]{{\em \textbf{#1}}}
\newcommand{\jl}[1]{{\em \color{red}{#1}}}
\def\multitd{{\it Multi3d}}
\def\Halpha{\mbox{H\hspace{0.1ex}$\alpha$}}
\def\Lyalpha{\mbox{Ly$\hspace{0.2ex}\alpha$}}
\def\Lybeta{\mbox{Ly$\hspace{0.2ex}\beta$}}
\def\feci{\mbox{\ion{Fe}{14}~211\AA}}
\def\fedi{\mbox{\ion{Fe}{12}~193\AA}}
\def\feni{\mbox{\ion{Fe}{9}~171\AA}}
\def\hedi{\mbox{\ion{He}{2}~304\AA}}
\def\fec{\mbox{\ion{Fe}{14}~274\AA}}
\def\fed{\mbox{\ion{Fe}{12}~195\AA}}
\def\fet{\mbox{\ion{Fe}{10}~184\AA}}
\def\fen{\mbox{\ion{Fe}{9}~188\AA}}
\def\sis{\mbox{\ion{Si}{7}~275\AA}}
\def\nc{\mbox{\ion{N}{5}~1238\AA}}
\def\cc{\mbox{\ion{C}{4}~195\AA}}
\def\os{\mbox{\ion{O}{6}~195\AA}}
\def\hed{\mbox{\ion{He}{2}~304\AA}}
\def\CaII{\mbox{\ion{Ca}{2}}}
\def\cair{\mbox{\ion{Ca}{2}~8542\AA}}
\def\CaIII{\mbox{\ion{Ca}{3}}}

\begin{abstract}

We have performed a 3D radiative MHD simulation of the solar atmosphere. 
 This simulation shows a jet-like feature that shows similarities to the type~II spicules 
observed for the first time with Hinode's Solar Optical Telescope. Rapid Blueshifted Events (RBE's)
on the solar disk are associated with these spicules. 
Observational results suggest they may contribute significantly in supplying 
the corona with hot plasma. We perform a detailed  comparison of the properties 
of the simulated jet with those of type~II spicules (observed with Hinode) and 
RBEs (with ground-based instruments). We analyze a wide variety of 
synthetic emission and absorption lines from the simulations including 
chromospheric (\ion{Ca}{2}~8542\AA , \ion{Ca}{2}~H  and \Halpha) to transition region and 
coronal temperatures (10,000 K to several million K). We compare their synthetic intensities, 
line profiles, Doppler shifts, line widths and asymmetries with observations 
from Hinode/SOT and EIS, SOHO/SUMER, the Swedish 1-meter Solar Telescope 
and SDO/AIA. Many properties of the synthetic observables resemble the observations, 
and we describe in detail the physical processes that lead to these observables. 
Detailed analysis of the synthetic observables provides insight into how observations
should be analyzed to derive information about physical variables in such a dynamic
event. For example, we find that line-of-sight superposition in the optically thin atmosphere
requires the combination of Doppler shifts and spectral line asymmetry to determine 
the velocity in the jet. In our simulated type~II spicule the lifetime of the asymmetry of the 
transition region lines is shorter than of the coronal lines. Other properties differ from 
the observations, especially in the chromospheric lines. The mass 
density of the part of the spicule with a chromospheric temperature is too low to 
produce significant opacity in chromospheric lines. The synthetic \cair\ and \Halpha\ 
profiles therefore do not show signal resembling RBEs. These and other 
discrepancies are described in detail, and we discuss which mechanisms and 
physical processes may need to be included in the MHD simulations to mimic 
the thermodynamic processes of the chromosphere and corona, in particular
to reproduce type~II spicules.  

\end{abstract}


\section{Introduction}

The \Halpha\ line, which reveals the relatively cool plasma in the chromosphere, 
shows that the upper chromosphere is dominated by highly dynamic spicules. Our view of 
spicules has been revolutionized thanks to advanced instruments such as the Hinode 
satellite \citep{2007SoPh..243....3K} and the Swedish 1-meter Solar Telescope (SST), 
where adaptive optics and image post-processing are  necessary \citep{van-Noort:2005uq}. 
Using Hinode observations, \citet{de-Pontieu:2007kl} distinguished at least two different 
types of spicules. 

The first type of spicules (so-called type~I) reach heights above the 
photosphere of 2-9~Mm, have a lifetime of 3-10 minutes and shows 
up- and downward motion \citep{Beckers:1968qe,Suematsu:1995lr}.
Type~I spicules are probably the counterpart of the dynamic fibrils on the disk and
they follow a parabolic path in space and time. This evolution 
is caused by a magneto-acoustic shock wave passing through the chromosphere 
\citep[][among others]{Shibata:1982qy,Shibata:1982fk,De-Pontieu:2004hq,
Hansteen+DePontieu2006,De-Pontieu:2007cr,Heggland:2007jt,Martinez-Sykora:2009kl,Matsumoto:2010lr}. 
Although the formation of type~I spicules seems to be well understood, even the most 
recent studies do not include various processes that may play an important role in
chromospheric dynamics and/or diagnostics such as time dependent ionization, 
generalized Ohm's law, 3D radiative transfer with scattering, etc). The impact of these 
various effects on type~I spicule evolution should be tested and compared with 
observations.
 
The second type of spicules (so-called type~II) reaches greater heights 
($\sim 6.5$~Mm) and has  shorter lifetimes ($\sim 100$s) than type~I 
spicules \citep{de-Pontieu:2007kl,Pereira:2012yq}. In addition, 
type~II spicules show apparent upward motions of order $50-100$~km~s$^{-1}$ 
and at the end of their life they usually suffer a rapid fading in images taken in 
chromospheric lines. The counterparts on the disk appear as rapidly moving 
absorbing features in the blue wing of chromospheric lines 
\citep{Langangen:2008fj,Rouppe-van-der-Voort:2009ul}. 
In contrast to type~I spicules, type~II spicules are not well understood. For example, 
some observations suggest the type~II spicules are impulsively and continuously 
accelerated while being heated to at least transition region temperatures 
\citep{De-Pontieu:2009fk,De-Pontieu:2011lr}. Recent observations \citep{Sekse:2012fk}
add complexity by indicating that some type~II spicules also show a decrease 
or a more complex velocity dependence with height. These spicules seem to 
show emission in transition region and coronal lines \citep{De-Pontieu:2011lr}; 
but \citet{Madjarska:2011fk} using lower signal-to-noise observations suggest that 
this emission comes from cold plasma. In addition, SDO/AIA observations show 
that some type~II spicules observed at the limb in the channel 304\AA\ (\hed) but it is unclear how common this is. The lack of detailed observational constraints poses significant challenges to theoretical modeling. As a result there is currently no model that can describe all aspects of type~II spicule formation.

Other motions, apart from the apparent upward motion, are observed in 
type~II spicules. For instance, these features show swaying motions at the limb 
of order $10-30$~km~s$^{-1}$, suggesting Alfv\'enic waves as observed 
in the chromosphere by \citet{De-Pontieu:2007bd} and \citet{Okamoto:2011kx}, 
and in the transition region and corona by \citet{McIntosh:2011fk}. 
\citet{Suematsu:2008zr} suggested that some spicules show multi-thread structure as 
result of a possible rotation. The torsional motions along spicules were 
suggested in earlier reports \citep{Beckers:1972ys}, but only recently 
unequivocally established using high resolution spectroscopy at the 
limb by \citet{De-Pontieu:2012bh}. The latter establish that type~II spicules 
show torsional motions of order $25-30$~km~s$^{-1}$. Similar torsional motions 
were observed in transition region lines in macrospicules and explosive events using 
both Dopplergrams and line profiles \citep{Curdt:2011uq}. 

As a result of the highly dynamic and finely structured nature of this type of spicules and
the complex physical processes governing the chromosphere (radiation, 
time-dependent ionization, partial ionization effects, etc.) our understanding of 
type~II spicules remains limited, as mentioned above. It is important that a 
successful model of these jets also explains the impact of type~II spicules on the 
various atmospheric layers (from the chromosphere to the lower corona). 
\citet{de-Pontieu:2007kl} and \citet{Sterling:2010vn} suggested that type~II 
spicules are driven by magnetic reconnection. Another possible candidate, also
more elaborated, was presented by \citet{Martinez-Sykora:2011uq} where the 
chromospheric plasma is ejected as a result of being squeezed by the magnetic 
tension resulting from flux emergence. The resulting jet shows evidence of both 
high velocities in the chromosphere and heating of plasma to coronal temperatures. 
On this basis, the jet was tentatively identified as a type~II spicule. However, 
a full and detailed comparison of the properties of various synthetic observables 
of the jet with observations of spicules and their counterparts has not yet been 
performed. 

In this paper we expand on the work done by \citet{Martinez-Sykora:2011uq} 
by comparing synthetic images and spectra of the simulated type~II spicule 
with observations from Hinode, Solar Dynamics Observatory (SDO) 
\citep{Lemen:2012uq},  the SST, and SOHO.  
The description of the code 
and the setup of the simulation used to simulate the candidate type~II spicule 
are detailed in Section~\ref{sec:equations}. In order to compare with observations 
we calculated synthetic observables as described in Section~\ref{sec:synthetic}.
The various observations and the data reduction are described in Section~\ref{sec:obs}. 
Section~\ref{sec:chr} details the synthetic chromospheric spectra and images in 
\ion{Ca}{2}~8542\AA\ and \Halpha\ and the comparison with observations. 
The synthetic transition region and coronal observables of the spicule are studied 
in Section~\ref{sec:euv}; they are compared 
with observations in Section~\ref{sec:compobs}. 
We finish with a discussion in Section~\ref{sec:conclusions}.

\section{Numerical Method and setup} 
\label{sec:equations} 

The MHD equations are solved in a 3D computational domain using the 
{\it Oslo Stagger Code} 
(OSC), which is the predecessor of the Bifrost code \citep{Gudiksen:2011qy}. 
The core of the numerical methods is the same and has been described also in 
detail by \citet{Dorch:1998db,Mackay+Galsgaard2001,paper1,Martinez-Sykora:2009rw}.
Essentially the code solves the partial spatial derivatives using a sixth order 
accurate method. In addition, a fifth order interpolation scheme is used if variables 
are needed at other locations than their defined position with respect to the grid. 
Time stepping of the
MHD equations is performed using a modified version of the third order predictor-corrector procedure 
detailed by \citet{Hyman1979}.  
Numerical noise is suppressed by adding a high-order artificial magnetic 
hyper-diffusion and -viscosity. The magnetic and viscous energy dissipated 
through the hyper-diffusive operator is self-consistently implemented into the 
energy conservative equation

From the photosphere to the lower chromosphere the radiative 
flux divergence is calculated by wavelength and angle integration. 
The opacities and emissivities are assumed to be isotropic. 
In order to solve the radiative transfer equation we assume four group 
mean Local Thermodynamic Equilibrium (LTE) opacities to cover 
the entire spectrum \citep{Nordlund1982}. The transfer equation is 
reformulated calculating a mean source function 
in each bin. In addition, an approximate coherent scattering and a contribution 
from the thermal emissivity are included in the source function. The 3D scattering problem
is solved following \citep{Skartlien2000}, i.e., it is solved by iteration using 
one-ray in the integral for the mean intensity.

For the middle and upper chromosphere, 
non-LTE radiative transfer losses from hydrogen continua, hydrogen lines, 
and lines from singly ionized calcium are calculated using \citep{Carlsson:2012uq} 
recipes based on 1D dynamical chromospheric model 
\citep{Carlsson:1992kl,Carlsson+Stein1994,Carlsson:1997tg,Carlsson:2002wl}. 

Optically thin radiative losses are calculated in the upper chromosphere, 
transition region and corona assuming the coronal approximation:  hydrogen, helium, carbon oxygen, 
neon and iron elements are included. Atomic data is obtained from the HAO spectral diagnostics package  \citep{HAO_Diaper1994}.

The thermal conduction along the magnetic field lines is treated by operator splitting. The 
the conductive operator is solved using the implicit Crank-Nicholson method and is iterated to convergence using a multi-grid solver  \citep{Gudiksen:2011qy}.

\subsection{Initial and boundary conditions}
\label{sec:condition}

\begin{figure*}
\includegraphics[width=0.95\textwidth]{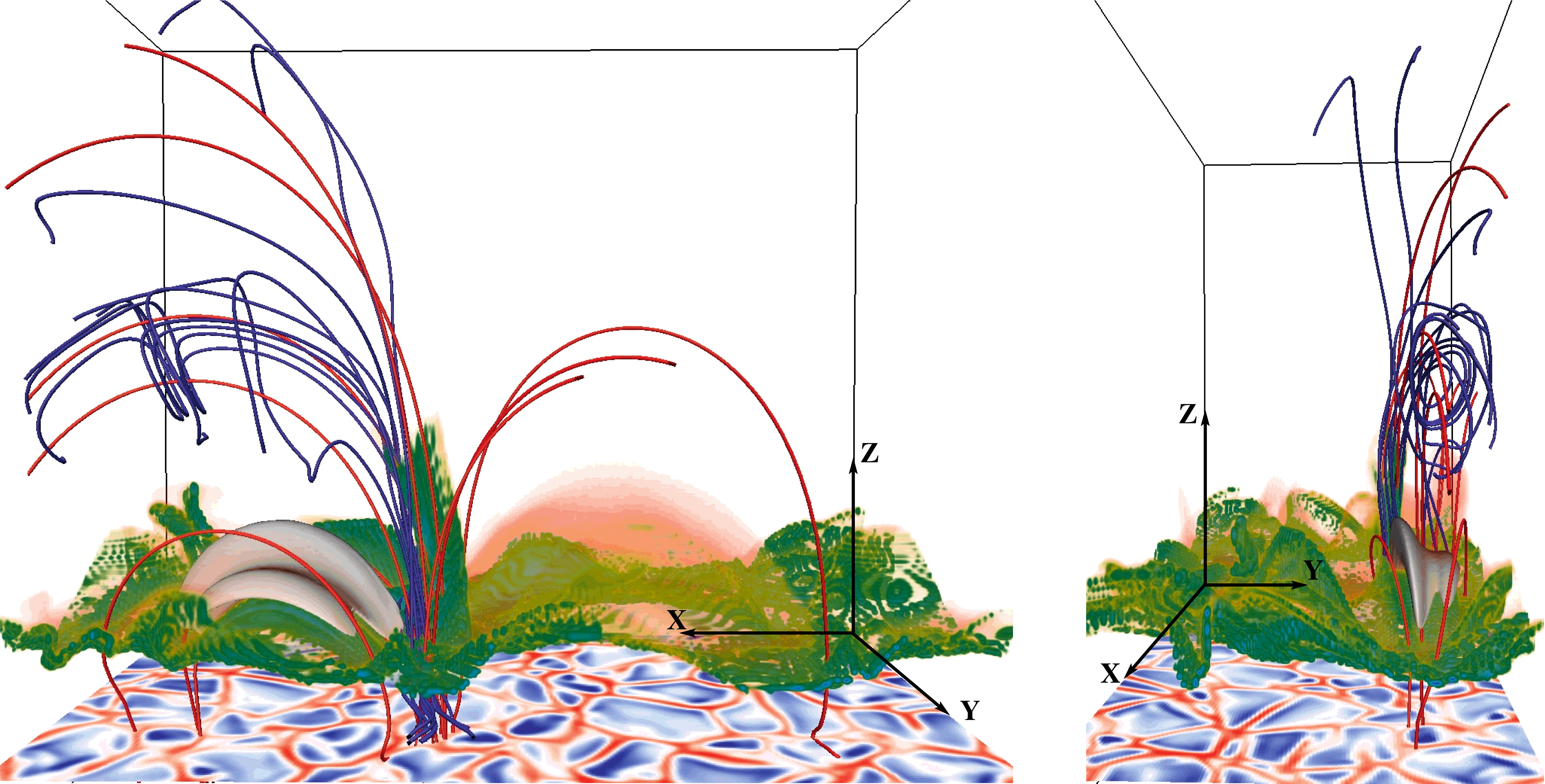}
\caption{\label{fig:init} Characterization of our radiation-MHD
  simulation. The panels show two 3D visualization of the
  computational domain at $t=1590$s. The vertical velocity in the
  photosphere is shown with a red-blue color scheme ([-2,2]~km~s$^{-1}$, blue are
  upflowing granules). Selected magnetic field lines are shown in red,
  selected velocity streamlines are blue. An isosurface of \feci\ emission
  is shown in grey. The \hedi\ emissivity is shown with the red-green
  semi-transparent color scheme with green corresponding to high emissivity.}
\end{figure*}

The simulation discussed in this paper was used to study flux 
emergence in \citet{paper1,Martinez-Sykora:2009rw}, 
type~I spicules in  \citet{Martinez-Sykora:2009kl} and 
the driving mechanism of a jet which resembles a 
type~II spicule in \citet{Martinez-Sykora:2011uq}. Here we calculate 
synthetic observables of this jet and describe the results. This simulation 
was named as ``B1'' in the previous work. 
The domain spans from the upper convection zone (1.5~Mm below the photosphere)
to the lower corona (14~Mm above the photosphere). The horizontal extent 
is $16$~Mm in the chosen $x$ direction and $8$~Mm in the 
$y$ direction, as shown in Figure~\ref{fig:init}. This figure shows a 
selection of magnetic field lines (red), velocity streamlines (blue), and the 
emissivity of \hedi\ (green-red semi-transparent color scheme) and 
\feci\ (grey-isosurface) at $t=1590$s.
The domain is evaluated in $256\times128\times160$ grid points using a uniform 
grid in the horizontal direction (i.e., the horizontal grid spacing is $65$~km) and 
a non-uniform grid in the vertical direction, where the spacing becomes larger at 
coronal heights as gradients are smaller and the scale heights larger than lower 
in the atmosphere (i.e., the minimal vertical grid spacing is $32$~km). After the 
transients from the initialization are relaxed, the simulation was run for one hour of 
solar time.

The initial mean magnetic intensity in the photosphere is $\sim 160$~G. 
The magnetic field is distributed in loop-shapes in the corona where
their footpoints are rooted in two bands along the $y$ axis in the photosphere 
($x \approx 7$~Mm and $x \approx 13$~Mm). As a result of this the magnetic
field lines in the corona are roughly oriented along the $x$-axis (see Figure~\ref{fig:init}). This figure corresponds to the instant $t=1590$s which 
shows the ejected type~II spicule as a vertical green finger in 
the left-hand panel. 

The trigger of the series of events that lead to the type~II 
spicule is a non-twisted flux tube that we introduced into the computational 
domain through its lower boundary, as detailed in Section~3.2 by 
\citet{paper1}. This horizontal flux tube is injected in a band along the $y$-axis
with a diameter of $1.5$~Mm at $x=8$~Mm. Note that the emerging 
field is mostly perpendicular to the orientation of the initial coronal loops
shown in Figure~\ref{fig:init}. The input parameters of the flux emergence 
are shown in \citet{Martinez-Sykora:2009rw}. 
The detailed description of how the flux emergence drives the type~II 
spicule is in \citet{Martinez-Sykora:2011uq}.

\subsection{Methods: Synthetic diagnostics}
\label{sec:synthetic}

To analyze the emergent emission of the simulated atmosphere, we calculate 
synthetic images and profiles of the chromospheric  \ion{Ca}{2}~8542\AA , 
\ion{Ca}{2}~H~3968\AA\ and \Halpha\ lines, which have been used to study 
rapid blueshifted events (RBEs) and type~II spicules with Hinode (\ion{Ca}{2}~H~3968\AA ) 
and SST \citep{Rouppe-van-der-Voort:2009ul,Sekse:2012fk}. 
In addition, we synthesized the emission in different EUV lines that form 
at transition region and coronal temperatures, including the dominant lines of the various 
SDO/AIA channels and various Hinode/EIS and SOHO/SUMER lines.

The \ion{Ca}{2}~8542\AA\  and \ion{Ca}{2}~H diagnostics have been calculated 
using the {\it RH} code 
\citep{Uitenbroek:2001dq}. An MPI-parallelized version of this code was used to 
solve the non-LTE radiative transfer problem in 1D on a column-by-column 
basis (1.5D approximation), for each snapshot of the simulation. 
\citet{Leenaarts:2009ly} find that a 1D treatment is a good approximation for 
Ca II H except at the very line core. Thus, its use is justified for wavelength-integrated 
Ca II filtergrams. A 5-level plus continuum model \ion{Ca}{2} atom was employed. 
Using the non-LTE opacity, source function, and radiation field, line profiles 
were calculated for the top and side view. For these calculations complete 
redistribution (CRD) was assumed. 

We computed the \Halpha\ line intensity with the \multitd\ code \citep{Leenaarts:2009ly}
in the same manner as done by \citet{Leenaarts:2012cr}. We employed a 5-level 
plus continuum hydrogen model atom and assumed statistical equilibrium. All lines 
were treated assuming complete redistribution. To mimic the effect of partial 
redistribution, the \Lyalpha\ and \Lybeta\ lines absorption profiles are assumed to be 
Gaussians with Doppler broadening only. The radiation field was computed in full 3D,
which is essential for the \Halpha\ radiative transfer, a 1.5D approach fails to 
reveal any chromospheric structure due to lack of lateral smoothing of the radiation 
field \citep{Leenaarts:2012cr}. Possibly important effects of non-equilibrium hydrogen ionization 
\citep{Leenaarts:2007sf,Leenaarts:2012cr} were not taken into account in the numerical 
simulations so we used LTE electron densities and temperatures.

Limb observations are synthesized assuming filter profiles from Hinode/SOT (\ion{Ca}{2}~H 3\AA\ FWHM). 

The emission for coronal EUV lines is calculated assuming the optically thin 
approximation under ionization equilibrium conditions. Hence, the synthetic 
frequency-integrated intensity in a spectral line is: 

\begin{eqnarray}
I(v,w) = \int_{l} A_{b}\, n_{\rm e}(v,w,l) \, n_{H}(v,w,l)\, G(T,n_{\rm e})dl, \label{eq:is}
\end{eqnarray}

\noindent where $l$, $v$, and $w$ are length along the line-of-sight (LOS), and the 
position in the surface perpendicular to the LOS, respectively. The synthetic images
result by integrating ($l$), e.g., along the $z$ axis, therefore $v$ and $w$ are in the 
$xy$ plane. Here $A_{b}$, $n_{\rm e}$, $n_{H}$, and $G(T,n_{\rm e}$)  
represent the abundance of the emitting element, the electron  
and the hydrogen densities, and the contribution function, respectively. 
The electron density is taken from the equation of state lookup table of the
simulation. We create a lookup table of the contribution 
function ($G(T,n_{\rm e})$) using the Solarsoft package for IDL 
{\tt ch\_synthetic.pro}, where the keyword GOFT is selected. Knowing the 
temperature ($T$), and the electron density ($n_{\rm e}$) for each grid-point, 
$G(T,n_{\rm e}$) is obtained by linear interpolation of the lookup table (in log space). To 
synthesize the plasma emission we use CHIANTI v.7.0 \citep{Dere:2009lr,Dere:2011kx}
with the ionization balance {\tt chianti.ioneq}, available in the CHIANTI
distribution. We synthesized observations for photospheric abundances
\citep{Grevesse:1998uq}.

The synthetic EUV spectral line profiles are computed 
\citep[as first done for a 3D simulation by][]{Peter:2006zk} following the methods 
described in \citet{Hansteen:2010uq} which assume that the lines are optically 
thin and in ionization equilibrium, as mentioned above. 

Once we have emergent spectral line profiles we use similar techniques as 
those used by observers to examine our results. We perform a single Gaussian fit to 
the profile to determine the Doppler shift and the line width. 
For the asymmetries of the line profiles, we use the Red-Blue asymmetry, hereafter 
RB asymmetry \citep{De-Pontieu:2009fk}. Briefly,  we fit a single Gaussian 
to the core of the line profile, and determine the line center of the Gaussian 
fit. The RB asymmetry is calculated by subtracting the integrated intensity 
within two spectral windows that are identical in size and located at identical 
wavelength offsets from the line center, with one window towards the red 
and one towards the blue \citep{Martinez-Sykora:2011fj}. 

Several studies have been done and prove the importance of the time dependent 
ionization for several of the EUV lines that we are using here 
\citep{Joselyn:1975vn,Mariska:1982uq,Hansteen:1993kx,Bradshaw:2009uq,Judge:2012uq,kosovare:2012}.  We 
postpone a detailed study of the (E)UV emission in type~II spicules taking into 
account time dependent ionization to a follow-up paper \citep[see][for 
an analysis of time dependence ionization in Bifrost simulations using a similar 
domain]{kosovare:2012}. Some of the results based on EUV lines presented here 
may thus be subject to the limitation of considering ionization equilibrium conditions.

\section{Observations}\label{sec:obs}

We perform a detailed comparison between these synthetic observables 
and solar 
observations. We use observations from the Swedish 1-m Solar Telescope 
\citep[SST,][]{Scharmer:2003ve} on La Palma using the CRisp Imaging SpectroPolarimeter 
 \citep[CRISP,][]{Scharmer:2008zv} instrument. We used 
\Halpha\ and \ion{Ca}{2}~8542\AA\ spectral line profiles obtained on 2012 
July 2 at 08:36-09:35 UT. The field of view was 
53\arcsec x53\arcsec\ with a spatial scale of 0.0592\arcsec/pixel.
The target was a network region close to disk center at solar
coordinates $(x, y) \sim (-15\arcsec, 50\arcsec)$ and $\mu=1$. 
The \ion{Ca}{2}~8542\AA\ line was sampled at 47 spectral positions from -2717~m\AA\  
to +2717~m\AA\, with a sampling of 110~m\AA\ in the line core region (sparser in the 
wings). The \Halpha\ line profile was sampled at 35 spectral positions in the blue wing 
from -2064~m\AA\ to 1290~m\AA\ with steps of 86~m\AA\ in the line core region (sparser 
in the wings). High spatial resolution was achieved with 
aid from adaptive optics and image restoration \citep{van-Noort:2005uq}. 

To study the impact of the type~II spicules in the 
transition region and corona we used the same data set as in 
\citet{De-Pontieu:2011lr} (see this reference for details). In short, the Hinode/SOT 
\Halpha\ (-868m\AA )  images were taken on 2010 April 25 from 01:55:38 to 02:55:02 
UTC with a cadence of 12s and a pixel size of 0.159\arcsec. The data were corrected for 
dark current, flatfielded, and co-aligned as a timeseries. The AIA/SDO images were 
taken at a cadence of 8s, with pixels of 0.6\arcsec, and exposure times twice as long
 as during normal AIA operations. To maximize signal to noise we made a 
new time series at a cadence of 16s where each image is the sum of two original images. 

Finally, for Doppler shifts and RB asymmetries of the transition region and coronal lines 
we used a combination of SOHO/SUMER  and Hinode/EIS observations. We used a sit-and-stare
sequence of a quiet Sun region at disk center obtained with SOHO/SUMER on 25 April 1996. 
These observations show a one hour long sequence of \ion{N}{5} 1238 \AA{} spectra 
that cover a quiet Sun region, including a network region at $y=-30$\arcsec\ to $y=-20$\arcsec . 
We use the same data set as \citet{Tian:2012uq} for the Hinode/EIS observations (see the 
references for details of the data set, calibration and the methods to calculate the 
Doppler shifts and RB asymmetries). Unfortunately, current observations do not 
include simultaneous coverage of clean spectral lines that are formed over a 
temperature range from the low transition region to the corona. 

\section{Results}
\label{sec:results}

\citet{Martinez-Sykora:2011uq} described a physical process of a jet 
which in several aspects resembles a type~II spicule as described in the 
current  paper. While some of the properties of this jet resemble type~II spicules, 
we only have a single example in our simulation. We discuss possible reasons 
below. In summary, this process follows naturally as a consequence of the 
constantly evolving magnetic and thermal environment of the modeled plasma.
The spicule is composed of chromospheric material that is rapidly ejected from 
the chromosphere into the corona, while being heated. The source of the ejection 
of the chromospheric material is located in a region with large field gradients 
and intense currents leading to strong Lorentz forces that squeeze the plasma. 
This increases the pressure gradient leading to strong upflows along the magnetic 
field lines. Most of the heating comes from magnetic energy dissipation which 
appears as a result of the interaction between the ambient field and the emerging 
fields that straightened as they expanded into the atmosphere. The resulting 
interaction between the emerging and ambient field lines leads to a tangential 
discontinuity and subsequently strong heating. This paper focuses on type~II 
spicules, for brevity referred to as ``spicules". When we refer to type~I spicules, 
it will be explicit.

\subsection{Synthetic Vs Observed Chromospheric Observables}~\label{sec:chr}

\begin{figure}
\includegraphics[width=0.45\textwidth]{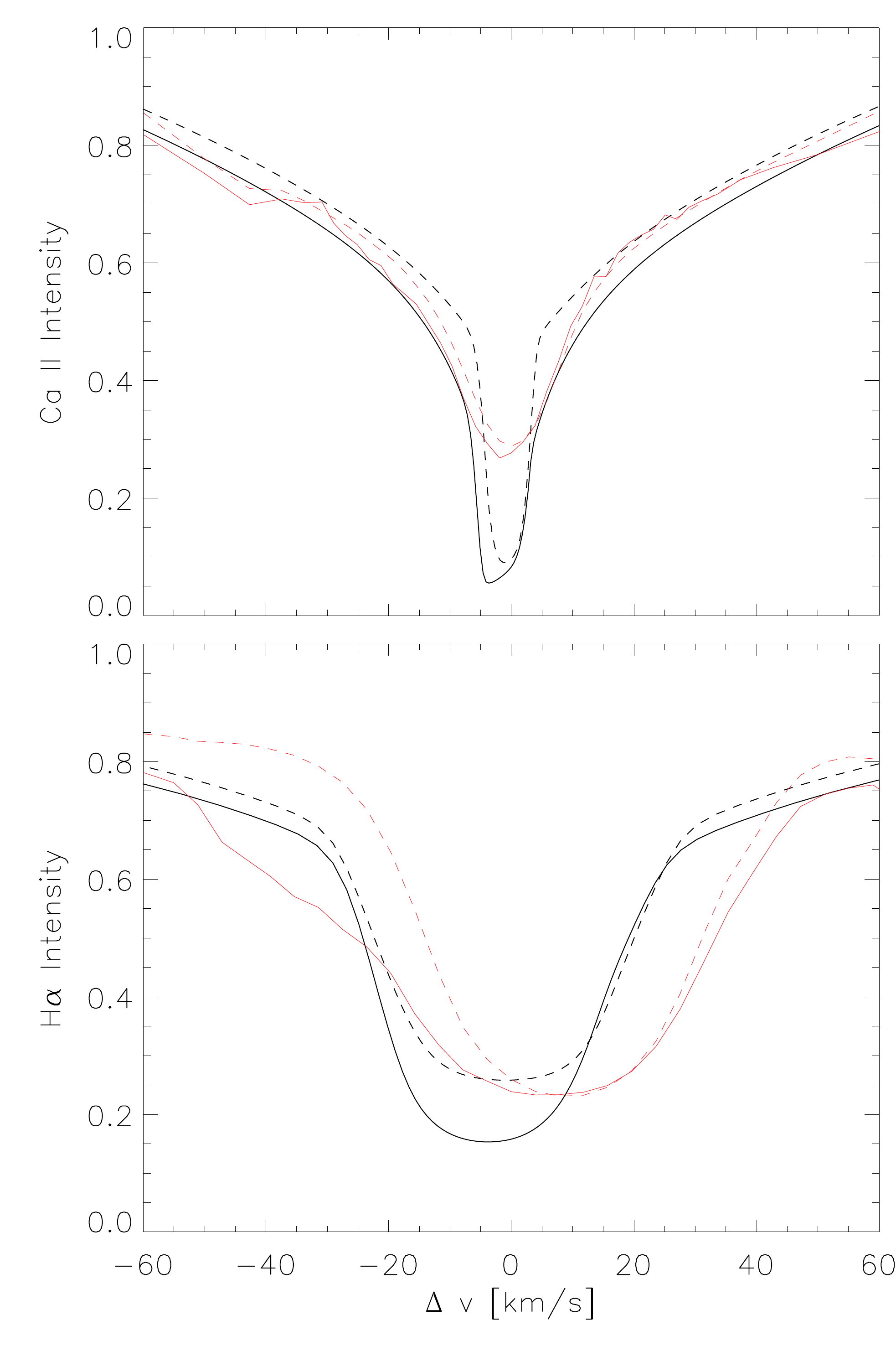}
\caption{\label{fig:cap} Comparison of observed on-disk RBEs (red
  curves) and the synthetic line profiles of the synthetic spicule
  (black curves, at simulation time $t=1590$s, at the location of the
  red cross in Figure~\ref{fig:breakca}). The top panel shows the
  \cair\ line, the bottom panel the \Halpha\ line. The solid curves are
  the line profile in the observed RBE and the synthetic spicule, the
  dashed curves are the average profile of a region surrounding the
  RBE and synthetic spicule. The intensities are normalized on the
  maximum intensity of the mean profile.}
\end{figure}

\citet{Martinez-Sykora:2011uq} described how chromospheric material is
ejected into the corona, but what does the jet look like in synthetic
chromospheric observations? In this subsection we show how the jet 
that resembles a type~II spicule appears in synthetic chromospheric 
diagnostics. Let us begin with discussing the on-disk view and then proceed 
to the off-limb view.

\subsubsection{On-disk appearance}~\label{sec:chrom-on-disk}

In observations on the disk, RBEs appear as a clearly separate component 
of absorption in the blue wing of the \cair\ and \Halpha\ profiles 
\citep[for example, see Figure 1 of][]{Rouppe-van-der-Voort:2009ul}. 
The solid red curves in Figure~\ref{fig:cap} show typical examples of RBE 
profiles, with a clearly lower-than-average intensity in the blue wing, 
with an unaffected red wing. In contrast, the synthetic profiles of the
spicule do not show such blue-wing asymmetry. Instead both the
\cair\ and \Halpha\ profiles show an asymmetric line core, whose
minimum intensity are blue-shifted $\sim 4$ and 
$\sim 5$~km~s$^{-1}$ respectively from the rest wavelength of the 
line. The shift is much smaller than the upflow velocities in the 
simulated spicule. Figure~\ref{fig:cap} is chosen at $t=1590$s, i.e., 
when the spicule reaches the longest insertion into the corona.

\begin{figure*}
\includegraphics[width=0.95\textwidth]{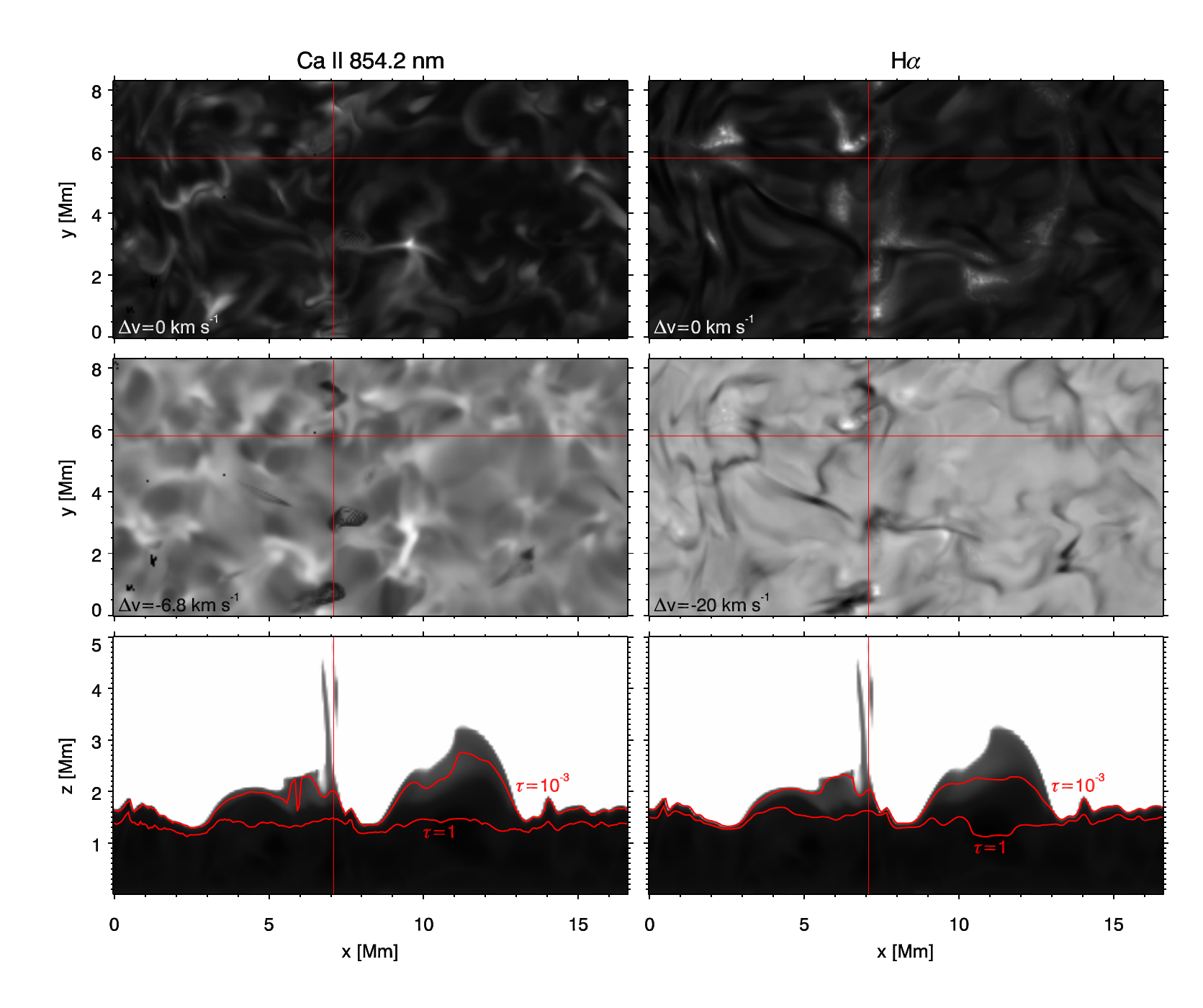}
\caption{\label{fig:breakca} Images of the on-disk \cair\ (left-hand
  panels) and \Halpha\ intensity (right-hand panels) at line-center
  (top row) and in the blue wing (middle row). The Doppler-shifts are
  indicated in the lower left corner of each panel. The red crosshairs
  indicate the location of the spicule and the line profile shown in
  Figure~\ref{fig:cap}. The bottom row shows an $xz$-slice of the
  temperature in the simulation (clipped at 20~kK) along the
  horizontal red line in the upper panels. The height of
  $\tau=10^{-3}$ and $\tau=1$ are shown as red curves. The vertical
  red line again indicates the location of the spicule. }
\end{figure*}
 
Figure~\ref{fig:breakca} shows why the synthetic profiles do not show
a blue-wing absorption coefficient. The top row shows line-center
images, the location of the spicule does not stand out (this is also the case in
observations). The middle row shows blue-wing images. Here the spicule
location shows up as small patches of below-average intensity, but it
does not stand out from the other dark structures in the image. The
bottom row finally shows why the spicule does not leave an imprint in
the profiles and images. The spicule is visible as the thin cold jet
protruding into the hotter coronal material. However, the overplotted
optical depth curves show that the spicule has a vertical optical
thickness smaller than $10^{-3}$, too small to leave an imprint
on the emergent profiles: the optical thickness at $z=3$~Mm is
actually only $7 \, 10^{-7}$. This lack of opacity can be caused by too
low mass density and/or too high temperature that lead to ionization to \CaIII\ and
therefore too little opacity. We tested the effect of the temperature by assuming
all calcium in the atmosphere is in the form of \CaII\ and computing
the upper bound of the optical thickness of the spicule given the mass
density. The spicule then gets an optical thickness at $z=3$~Mm of
0.02 which is also too small to leave an imprint on the
emergent profile. Having a lower temperature spicule (or having a
delayed ionization of \CaII\ through non-equilibrium ionization)
would thus not be enough to get an observable signal; the mass
density is also too low by at least a factor of five.
This is consistent with the low electron
density in the simulated spicule ($\sim2 \, 10^{9}$~cm$^{-3}$ above $z=3$~Mm)
compared to electron densities derived from spicule observations
\citep[$5 \, 10^{10}$---$2 \, 10^{11}$~cm$^{-3}$][]{Beckers:1968qe,Beckers:1972ys}.

The lack of density may be caused by the fact that the spicule does not eject 
enough chromospheric material.  It may also be
that the spicule expands too fast in the corona due to the magnetic
field expansion. The simulations have a rather simplified magnetic field 
configuration (Figure~\ref{fig:init}) compared to those on the Sun, e.g., the 
simulated emerging flux tube is a longitudinal flux tube along the $x$-axis. 
In contrast, in the Sun the small-scale emergence occurs frequently, and 
the orientation of these events is random \cite{Martinez-Gonzalez:2007vn}. 
The observed field strengths are also different from those in our simulation. 
In addition, the modeled ambient magnetic field does not show the same salt 
and pepper distribution as in the solar photosphere. The simulated magnetic 
field distribution is most likely also not reproducing the interaction between 
network and internetwork and the various scales observed in the Sun. Note 
that the magnetic complexity seems to play a role in eruptions 
\citep[see, e.g., ][]{Georgoulis:2012fk} and it might be expected that a more 
complicated modeled field could lead to more violent chromospheric and 
coronal dynamics. This may explain why these jets are 
rare in the simulations, i.e., because we lack magnetic field complexity and 
the interaction between various magnetic features

The core of the synthetic profiles are narrower than the observed profiles 
(Figure~\ref{fig:cap}). This discrepancy between synthetic and observed 
profiles exists not only in the spicule but in all current simulations. 
Recent analyses suggest that this is related to the relatively 
low spatial resolution of the simulations. 

\begin{figure}
\includegraphics[width=0.49\textwidth]{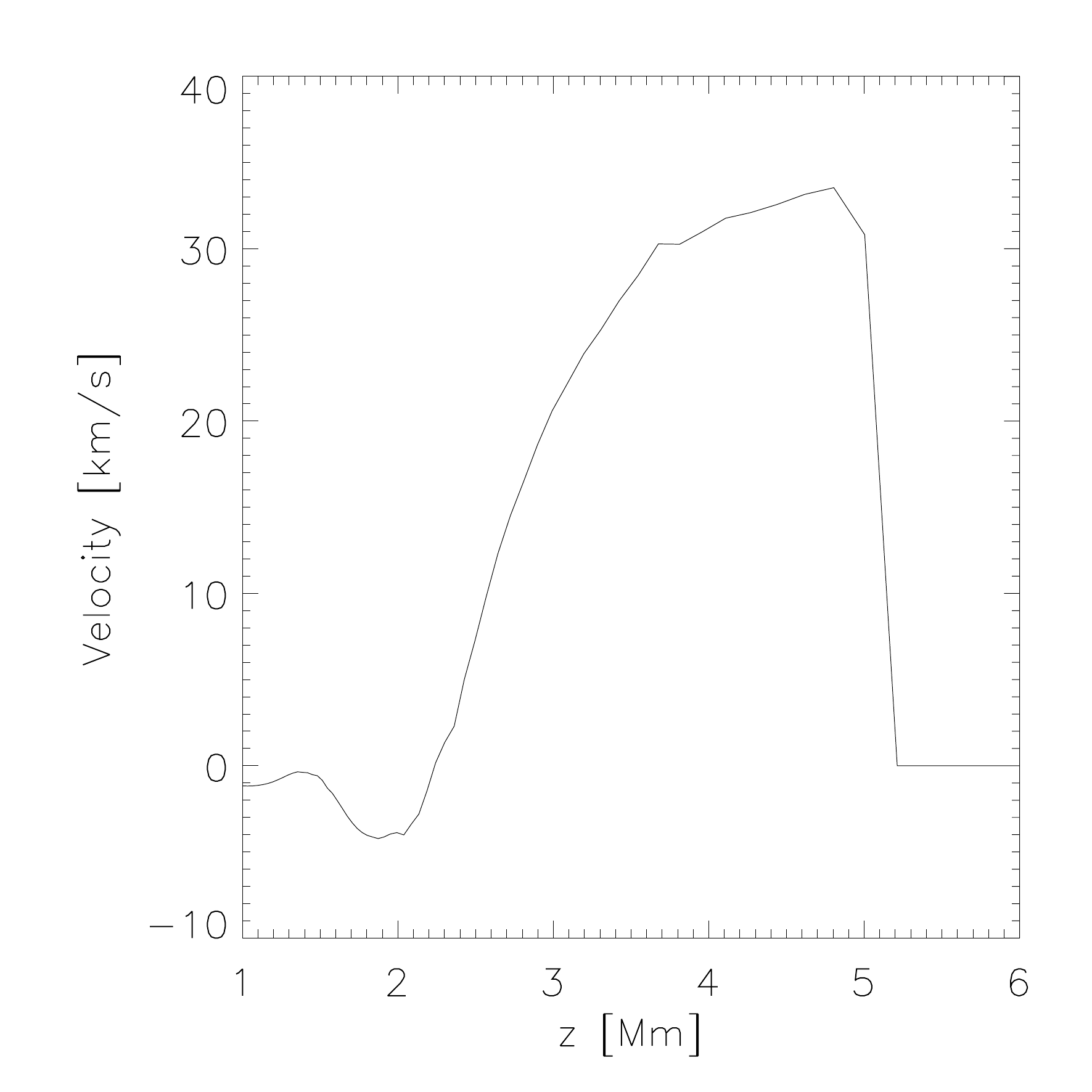}
\caption{\label{fig:provel} Increase of projected velocity along the simulated spicule. 
	 The horizontal mean of the velocity projected at an angle 
	of 30 degrees with respect to the vertical axis is shown as a function of height at 
	$t=1590$s where the spicule is located and the temperature is lower 
	than $4\, 10^{4}$~K. This increase is similar to observations which show an 
	increase of the Doppler shift with distance along the RBEs.}
\end{figure}

In contrast to the observations, where the spicules are usually not aligned with the 
LOS, the synthetic spicule is aligned with the vertical axis.
First results done by \citet{Rouppe-van-der-Voort:2009ul} revealed that most of the 
observed RBEs show an acceleration along them. Most recent and more detailed 
studies showed that observed RBEs can show a Doppler shift increase, 
decrease or a variable distribution with height \citep{Sekse:2012fk}. 
In our simulated spicule, we investigate how
the spicule aligned flows depend on height by calculating the velocity along a LOS
that is 30 degrees off the vertical axis.  Figure~\ref{fig:provel} shows this 
projected velocity at $t=1590$s for positions where 
the spicule is located and the temperature is lower than $4\, 10^{4}$~K. 
The projected velocity increases along the spicule, basically because the plasma is 
accelerated along the whole spicule and not at a single location. This acceleration is due to 
the fact that the compression acts along a large stretch along the spicule 
\citep{Martinez-Sykora:2011uq}. This may explain the observed increase of Doppler 
shift with height along RBEs  \citep{Sekse:2012fk}. 

\subsubsection{Off-limb appearance}

\begin{figure*}
\includegraphics[width=0.95\textwidth]{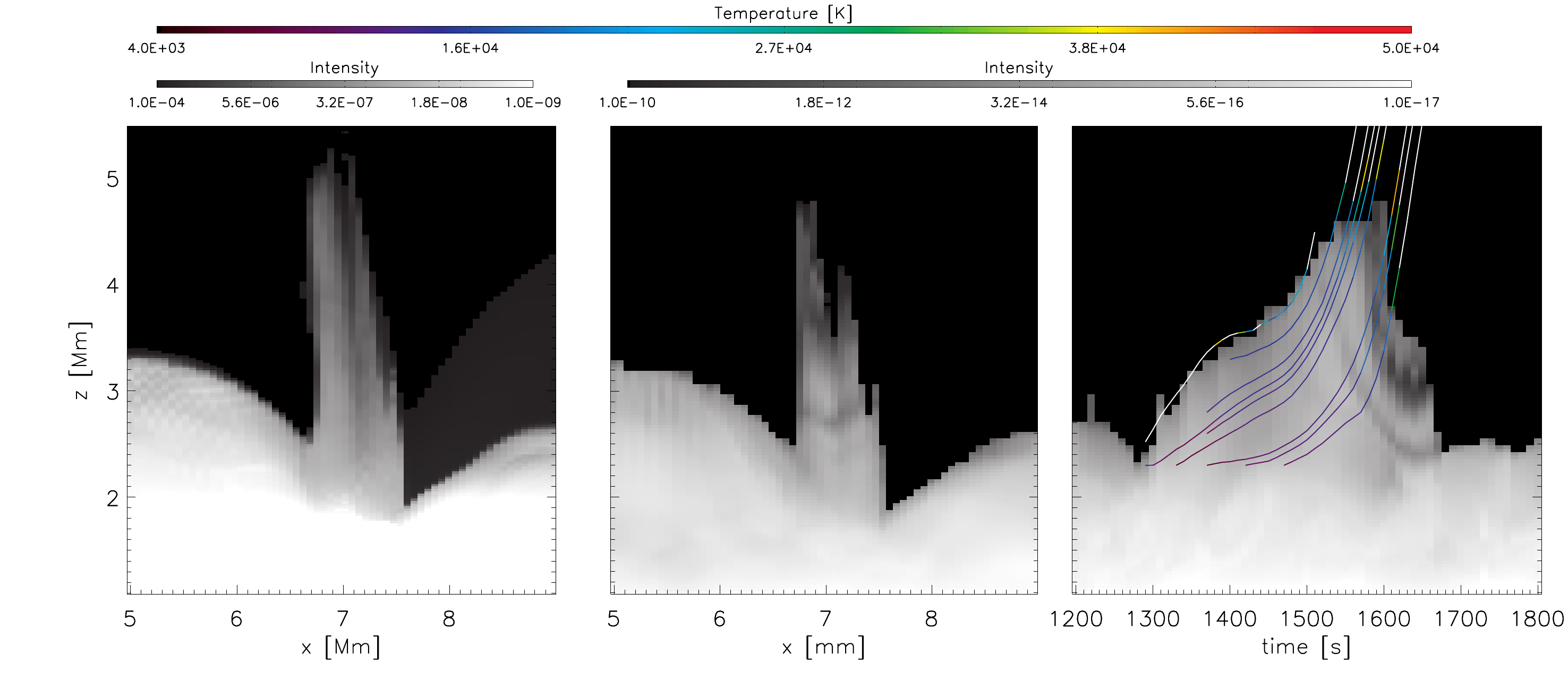}
\caption{\label{fig:caxz} Appearance of the synthetic spicule in
  chromospheric lines at the limb. \Halpha\ and \ion{Ca}{2}~H~3968\AA\ intensity from side-view 
	(left and middle panels respectively) at $t=1590$s and \ion{Ca}{2}~H intensity
	and as a function of time and height at $x=[6.7]\pm 0.2$~Mm
        (right panel). The intensity is shown on a logarithmic
        scale. The right panel also displays a number of test particle
trajectories to indicate the matter flow. The trajectories are color
coded with the gas temperature.}
\end{figure*}

Figure~\ref{fig:caxz} shows images of the simulated off-limb intensity at the
center of \Halpha\ (left), \ion{Ca}{2}~H~3968\AA\ (middle) as well as a $zt$-slice 
of \ion{Ca}{2}~H (right). The intensity was computed by solving the transfer equation 
along the $y$-axis for each $xz$ point in the simulation box. The effect of the 
curvature of the Sun was not taken into account. The spicule reaches 
heights up to 4.9~Mm for \ion{Ca}{2}~H and 5.5~Mm for \Halpha\ above 
the photosphere, i.e., this is a small type~II spicule but still within their observed 
height range \citep{Pereira:2012yq}. Note that in the spicule the synthetic \Halpha\ 
emission extends to a larger heights ($\sim0.6$~Mm) than the \ion{Ca}{2}~H 
emission. Even though this is at the limb and for a different \ion{Ca}{2} line,
it seems compatible with recent observations where the location of 
\cair\ RBEs are found closer to the footpoints than \Halpha\ RBEs
\citep{Rouppe-van-der-Voort:2009ul,Sekse:2012fk}.
The synthetic spicule also shows substructure suggestive of multiple threads
consistent with some observations of off-limb spicules with Hinode
\citep{Suematsu:2008zr}.  This structure is caused by the density and 
temperature variation within the spicule. This variation is caused by two 
factors: 1) the energy release is not uniformly distributed across and along 
the spicule; 2) the ejected plasma does not uniformly expand into the corona 
because the magnetic field lines do not uniformly expand with height.

One of the major discrepancies between the off-limb observations and synthetic observables 
is the difference in scale height intensity for the line core emission of \ion{Ca}{2}~H 
and \Halpha. At $2.2$~Mm above the limb of the quiet Sun, \citet{Pereira:2012yq} find a scale 
height of $2-3$~Mm in the \ion{Ca}{2}~H line. In contrast, in the synthetic 
observables in the vicinity and inside of our spicule event, we find intensity scale heights of only 
$0.25$~Mm for the \ion{Ca}{2}~H line and $0.6$~Mm for \Halpha . Further from 
the spicule, in regions where the atmosphere is not affected by flux emergence 
and the spicule event, the intensity scale heights are even smaller; $0.15$~Mm and $0.25$~Mm
 for \ion{Ca}{2}~H and \Halpha\ respectively \citep[the presence of flux emergence 
increases the intensity scale height as suggested
by][]{Martinez-Sykora:2009rw}.

As a result, the optical thickness at line center of the 
\Halpha\ spicule at $z=4$~Mm is $3\,10^{-5}$, i.e., this signal may be too faint to detect in 
real observations. We speculate that there are several reasons for the difference in 
intensity scale height: 
\begin{enumerate}[(a)]
\item As a result of the small size of the numerical domain the simulation 
lacks the LOS superposition of background intensity at spicular heights. 
\item The simulation likely has a spatial resolution that is too low.
\item As mentioned in
Section~\ref{sec:chrom-on-disk}, the spicule does 
not eject enough chromospheric material and/or expands too quickly into the corona. 
\item As alluded, the simulation does not mimic emergence of small-scale
flux on the Sun, i.e., 
the salt and pepper distribution and the small-scale flux emergence 
distribution in the simulation do not reproduce the observations well. 
\item Finally, current calculations do not take into account time-dependent 
ionization. The latter effect can in principle lead to  
larger formation heights for both \ion{Ca}{2}~H and \Halpha\ 
\citep{Wedemeyer-Bohm:2011oq,Leenaarts:2007sf}.
\end{enumerate}

At the limb the spicule shown in \ion{Ca}{2}~H (right panel of
Figure~\ref{fig:caxz}) rises $1$~Mm in $100$s, i.e., the apparent
upward speed is around $10$~km~s$^{-1}$.  This speed is significantly
smaller than the observed apparent velocities in \ion{Ca}{2}~H
filtergrams at the limb \citep{Pereira:2012yq}. In the early state of the 
spicule, the leftmost 
streamline indicates that spicular material at the top of the spicule
rises with the same speed as the apparent upward motion. Spicular
material that is accelerated upward at later times reaches larger
velocities, up to 40 km~s$^{-1}$. The apparent upward motion of the off-limb
emission is stopped as the gas in the spicule is heated to
temperatures above 20~kK and all calcium is ionized to
\ion{Ca}{3}. Therefore, the apparent acceleration in the second half 
of its evolution is limited by the ionization front. 
This process is similar to what \citet{Heristchi:1992kx} suggested. 
The spicule then fades rapidly from view in \ion{Ca}{2}~H
intensity. The spicule does not vanish because the plasma descends but
because Calcium is ionized away.

\subsubsection{Heating of the spicule out of the chromospheric line passbands}

\begin{figure*}
\includegraphics[width=0.95\textwidth]{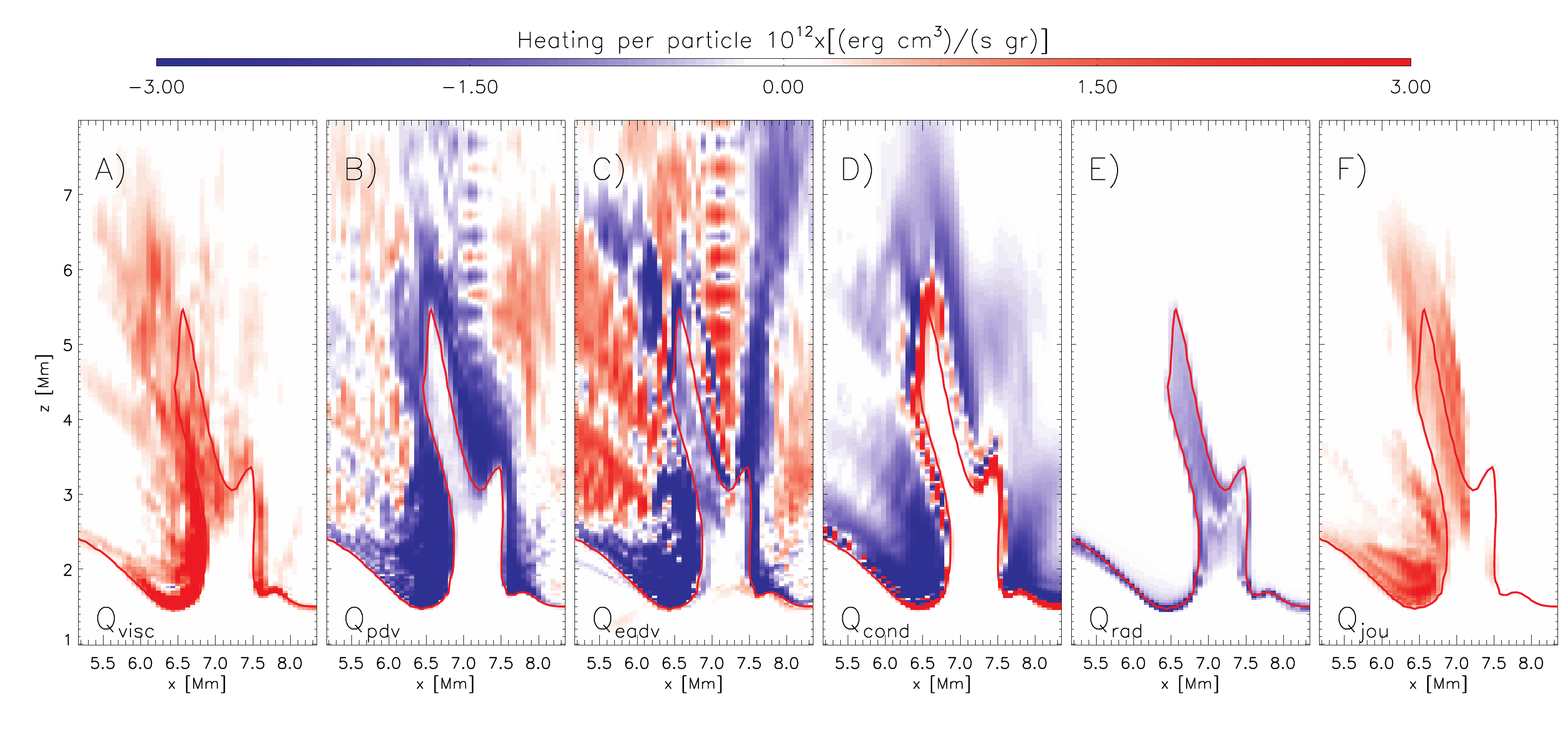}
\caption{\label{fig:joucond}  The various heating and cooling sources are shown in the 
	top row, i.e., viscous heating (Panel A), compression work 
	($P\nabla u$) (panel B), advection (panel C), conduction  (panel D), the 
	radiative term  (panel E) and the joule heating  (panel F). 
	The vertical cut  is at $y=6.175$~Mm, at $t=1590$s. 
	The temperature at $2\, 10^{5}$~K  is overlaid on the plots with red contours.}
\end{figure*}

Figure~\ref{fig:joucond} shows the 
various heating and cooling sources per particle in the spicule. The dominant heating 
mechanism is the Joule heating. However, the complexity of the spicule shows that 
other sources also can play a role in different parts of the spicule. 
For example, thermal conduction seems to be rather important in regions where the 
temperature of the plasma reaches $\sim 10^{5}$~K. These regions are located at 
the edge of the chromospheric material; the upward flowing plasma that stays cool the longest 
and that penetrates into the corona is finally heated by conduction (this 
parcel of plasma can be appreciated in the running time difference of \hedi\ 
shown in Figure~\ref{fig:intt} around $z\sim6$~Mm and $t=1650$s). 

Inside the chromospheric material of the spicule, the radiative cooling (Panel E in 
Figure~\ref{fig:joucond}) plays an important role counteracting the viscous and Joule 
heating (Panel A and F). The radiative cooling is greater in the earlier states of the 
spicule. When the spicule expands into the corona, the density drops, and the Joule 
heating also becomes more important: The chromospheric material heats up with 
time and radiative losses decrease. The largest contributions to the radiative cooling 
come from Hydrogen and Calcium ions.

Finally, one of the hot footpoints is heated mainly due to Joule heating 
(Figure~\ref{fig:joucond}). Afterwards, 
conduction (Panel D) and advection (Panel C) spread this heat along field lines. 
As can be appreciated in the 
figure, the regions where the different heating sources are concentrated are not 
easily linked, and the 3D structures are rather complex. We would like to remark 
that, in contrast, type~I spicules do not have strong sources of Joule heating, and 
the plasma suffers an expansion followed by a contraction 
\citep{Hansteen+DePontieu2006,Martinez-Sykora:2009kl}. Therefore, this type of 
spicules shows the well-known parabolic profile. However, the type~II spicule 
candidate shows a complex contribution of various heating and cooling terms, 
emphasizing the role of Joule heating. In addition, the chromospheric plasma 
does not experience an expansion followed by a compression, but it is ejected 
into the corona and heated. 

We are neglecting physical processes in the chromosphere that may
impact the evolution of the spicule: 
\begin{itemize}
\item The ion-neutral effects such as a
proper treatment of Generalized Ohm's law play an important role that
may change the way the spicule is heated \citep[see a discussion of
the importance of these effects in][]{Martinez-Sykora:2012fk}. 
\item Joule heating in the model is dependent on numerical
dissipation in order to keep structures resolvable on the numerical
grid, therefore they will occur on much larger scales than those found
on the Sun. This may have important implications for the details of
type II spicule heating. 
\item Ionization of hydrogen and helium is
treated with the simplifying assumption of LTE. Proper inclusion of
non-equilibrium ionization \citep{Leenaarts:2007sf,Leenaarts:2011qy} 
will lead to a different response of the thermodynamic state of the 
atmosphere to sudden heating.
\end{itemize}

\subsection{Emission in EUV, from TR to coronal lines: Impact into the corona} \label{sec:euv}

We study the impact of the simulated type~II spicule on the corona using synthetic EUV 
intensities and spectra under the optically thin approximation and ionization 
equilibrium conditions as described in Section~\ref{sec:synthetic}. 

\subsubsection{Intensity: Heating and cooling information}

\begin{figure*}
\includegraphics[width=0.99\textwidth]{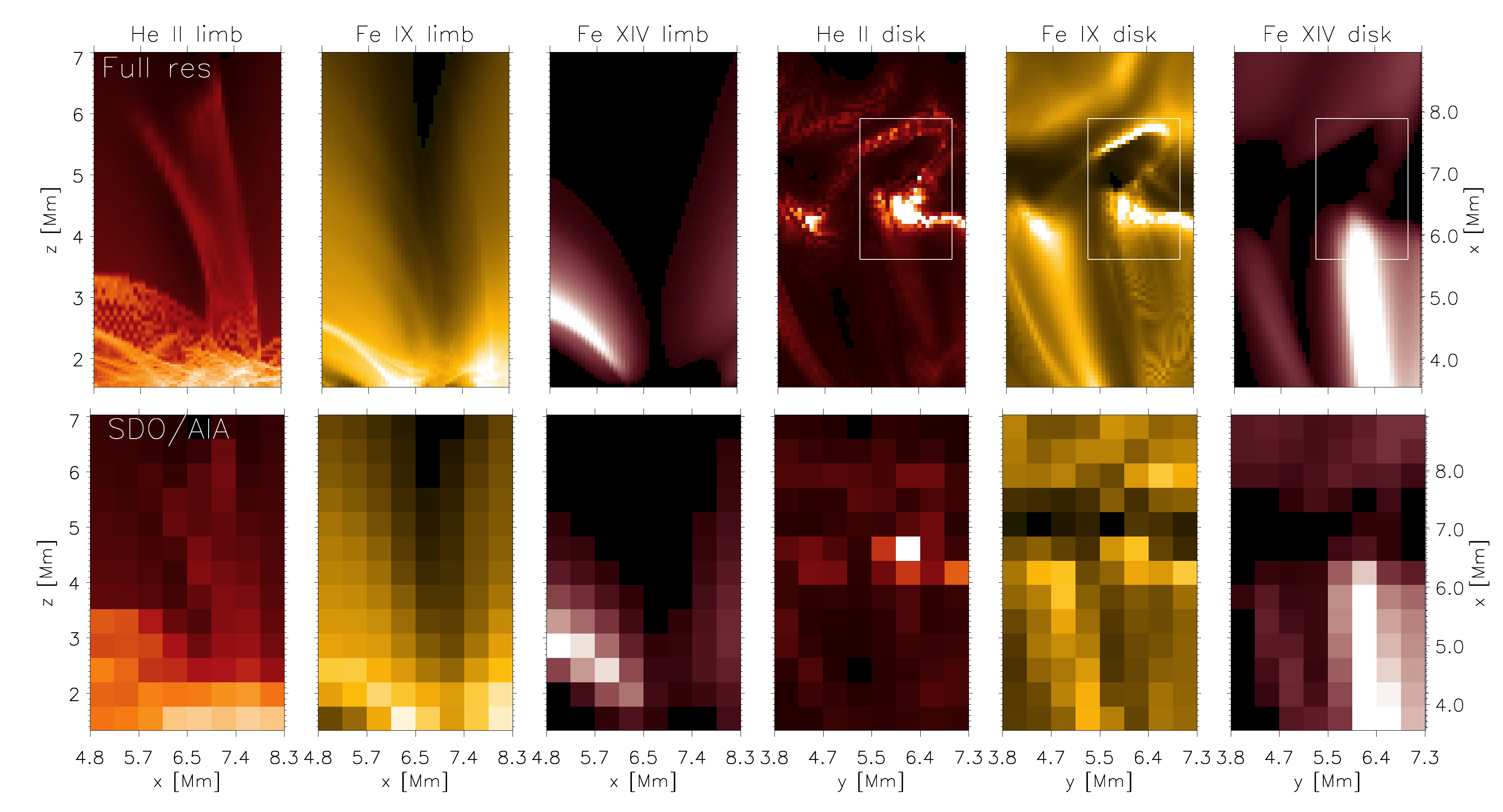}
\caption{\label{fig:int} Synthetic intensity maps at time $t=1630$s for \hedi\ (first and fourth 
	columns), \feni\ (second and fifth columns), and \feci\ (third and sixth columns) 
	for a limb view (first three columns) and top view (last three 
	columns). The top row is at the spatial resolution of the simulation and the 
	bottom row is taking into account the SDO/AIA spatial resolution.} 
\end{figure*}

During the evolution of the spicule, it shows considerable emission in transition 
region and coronal EUV lines in several different ions. Figure~\ref{fig:int} shows 
the intensity from the top and limb view coming from the ions \hedi , \feni , and \feci\ 
using full numerical spatial resolution (top) nd convolved to a Gaussian with the 
same width as the SDO/AIA PSF and rebinned to its spatial resolution (bottom 
panels). These lines constitute the strongest contributions to the 
304\AA , 171\AA\ and 211\AA\ SDO/AIA channels respectively \citep[the same 
channels were analyzed by][to show the emission of the type~II spicules in the 
corona]{De-Pontieu:2011lr}. We consider only the dominant ions to isolate the 
information of the specific SDO/AIA channels from the emission caused by other 
``non-dominant'' lines \citep{Martinez-Sykora:2011vn}. It is important to mention
that \hed\ line formation is poorly understood, so it is not clear that the coronal approximation 
approach is valid \citep{Feldman:2010lr}. Nevertheless, we calculated the line 
assuming coronal approximation and that the line is optically thin. 

It is interesting that the locations of the various emissions are not co-spatial 
due to the complex thermal process occurring inside and surroundings the 
spicule. For instance, the \hedi\ emission at the limb shows clearly the 
structure of the spicule, i.e., the shape of the ejected chromospheric material, 
but most of the emission coming from \feni\ is concentrated at the 
footpoint of the spicule, and for \feci\ it is also along field lines that 
connect to the footpoints of the spicule where most of the heating source is 
located. 

In the disk view, the emission coming from \hedi\  is not concentrated where 
the spicule is, but at the side boundaries of the ejected chromospheric 
material, rather similar to \feni . This is basically because at the footpoints 
of the spicule the density is higher than at the top of the spicule, 
and both the \hedi\ and \feni\ contribution functions peak near 
the transition region, i.e., surrounding the spicule. The spicule is heated but 
the plasma inside the spicule is never ``filled in'' 
with the emission of these lines because the spicule is not heated uniformly.
In contrast, the emission coming from \feci\ is located along the field lines 
that connect to the footpoint of the spicule. As a result, it is a significant challenge to 
link the emission coming from the different ions in space and time. 
Note that some of the detailed spatial structuring of the spicules observed 
in the different lines vanishes when taking into account SDO/AIA spatial resolution.

Disentangling emission from various ions in the spicule is difficult because of 
the complexity of the simulated jet. An additional complication, arises 
from the fact that the various SDO/AIA channels have contributions from other, typically 
non dominant lines 
\citep[][studied these contributions in 3D MHD simulations]{Martinez-Sykora:2011vn}. 
We find that the non-dominant lines do not play a significant role in the emission 
of the simulated spicule. We note that the most significant non-dominant lines in 
the 171, 193, and 211\AA\ SDO/AIA channels are emitted by ions formed at very similar low
transition region temperatures (e.g., \ion{O}{5}/VI). This means that significant 
emission from such ions in spicules would occur at the same location and time
for the various SDO/AIA channels. We can exploit this to distinguish between 
emission coming from the 
dominant ion and the non-dominant ions. As we can see in  
Section~\ref{sec:compobs} our model agrees well with the observed 
offset space and time in the SDO/AIA channels \citep{De-Pontieu:2011lr}.  
This supports the hypothesis that the spicule emission in the various 
SDO/AIA channels is not a low temperature contamination as suggested by 
\citet{Madjarska:2011fk}. 

\begin{figure*}
\includegraphics[width=0.95\textwidth]{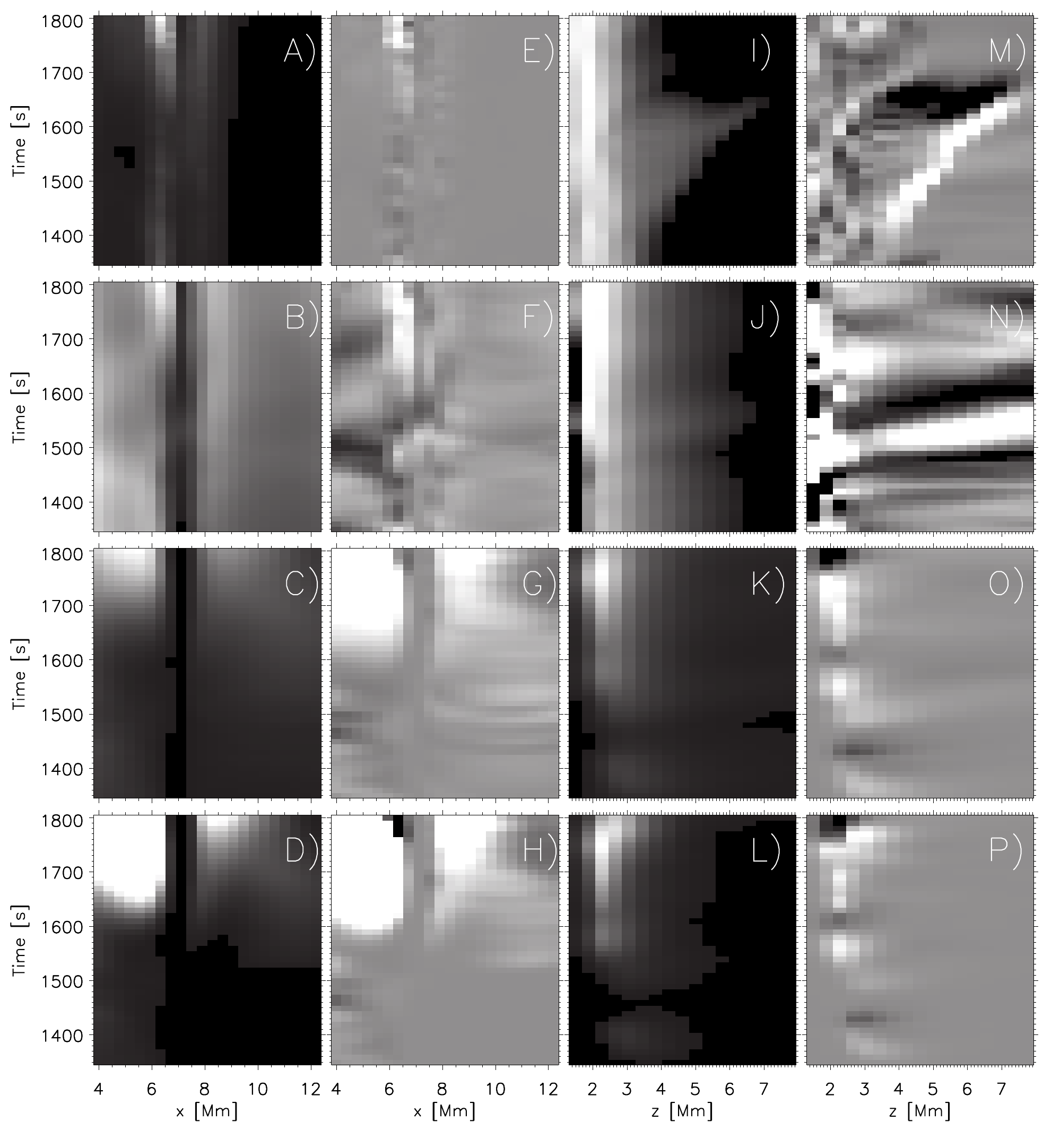}
\caption{\label{fig:intt} Intensity from top-view (two left columns) and limb 
	view (two right columns) using SDO/AIA spatial resolution (0.6 arcsec)
	as a function of time and length coming from \hedi\ (top row), 
	\feni\ (second row), \fedi\ (third row) and \feci\ (last row).
	The time differences are shown in the second and last column which are 
	calculated by subtracting the intensity of the two previous snapshots ($10$ and $20$s 
	earlier). For the intensity we integrated over 3 pixels in the $y$-direction 
	for the on-disk observations and in the $x$-direction for the limb observations.}
\end{figure*}

Figure~\ref{fig:intt} shows the intensity from the top-view (left two columns) and 
a limb view (right two columns) as a function of time and length along the spicule coming from 
\hedi\  (top row), \feni\ (second row), \fedi\ (third row) and \feci\ (last row). 
The running differences are shown in the second and last column. 
These lines correspond to the strongest contribution of the 304, 171, 
193 and 211\AA\ SDO/AIA channels respectively. The images 
on the disk (two left columns) show that all channels show strong emission for 
a relatively long time period. In fact, the emission increases in time over more than 
5 minutes. This increase starts earlier in cooler lines such as  \hedi\ and \feni . 
In addition, a weak brightening on the disk for the various filters  
is formed in the spicule when it is ejected ($t \sim 1550$s), 
but it is very faint and it can be appreciated only in the running difference. 
Additionally, the increase is fainter for hotter lines (compare with observations, Section~\ref{sec:compobs}). 
It is also interesting that on the disk \hedi\ is concentrated in a small 
region, but for the coronal lines the emission is spread out over more 
than $2$~Mm. Note that the various plots in Figure~\ref{fig:intt} do not 
follow a loop but are straight lines.  In fact for \feni , \fedi , and \feci\ the 
source of this strong emission comes from the footpoint of the spicule which 
is heated by Joule heating (see at $x=6.2$ and $z=2.2$~Mm in 
Figure~\ref{fig:joucond}) and expanded along the field lines by conduction. 

At the limb, a weak signal can be seen also in \fedi\ and \feci\ after the 
spicule is ejected. The type~II spicule candidate starts to appear at the side 
view around $t=1400$s. The coronal lines seems to agree with the 
observations of \citet{De-Pontieu:2011lr}. The emission moves 
into the corona as a propagating coronal disturbance at  $\sim100$~km~s$^{-1}$. 
Note that the plasma does not travel at that speed, i.e., the apparent propagation 
of the emission is a combination of mass flow, waves, and thermal conduction 
front along the field lines. The plasma flow is sonic up to temperatures $8\, 10^5$~K 
and subsonic at higher temperatures \citep{Martinez-Sykora:2011uq}. The 
strongest signal is from \feni , but is also noticeable in  \fedi\ and 
\feci. Nevertheless, the strongest contribution of these two lines comes roughly  
one hundred seconds later. In a similar fashion as seen on the disk,  for \feni , \fedi , and 
\feci\ the source of this strong emission comes from the footpoint of the spicule, 
which is heated by Joule heating and propagated along the field lines by 
conduction into the upper layers in the corona. In contrast, the formation 
temperature of \hedi\ is too low for the conduction. However, thermal conduction 
plays a role in the fading in the \hedi\ emission around $t=1650$s: as the last 
remaining chromospheric parcel penetrates into the corona, the plasma around 
$10^{5}$~K is heated beyond the \hedi\ formation temperature (see panel M around 
$z\sim 6$~Mm and $t=1650$s in Figure~\ref{fig:intt}). We note the lifetime of the 
profiles in \hedi\ is of the same order as in the observations (a few hundred seconds). 
Compare these results with the observations described in Section~\ref{sec:compobs}. 

It is important to mention that the length, ``propagation'' and duration of the 
emission of \feni , \fedi , and \feci\ are strongly dependent on the magnetic 
field configuration. In fact, our domain is small and most of the magnetic field 
lines are confined within the small domain with only a few open field lines that 
cross the top boundary. Plasma on the latter loses heat because the upper 
boundary is open. As a result of this, the open field lines do not maintain the 
high temperatures because the energy is transported through the top of the 
domain due to thermal conduction. In contrast, the closed field lines produce 
small hot loops that are heated by the Joule heating coming from the footpoint 
of the spicule. This magnetic field configuration differs from that usually found 
on the Sun (Section~\ref{sec:compobs}). 
The field lines in the Sun are probably more complex and, e.g., in plage regions
the magnetic field lines connect over very large distances, so that heat can propagate 
further and be distributed to greater distances. In a similar manner, the magnetic field 
configuration of our simulation is also completely different from the magnetic 
field in coronal holes. The field configuration of the 
simulation, despite the simplicity of the magnetic field configuration and 
the small size which forces it to have small loop structures in the corona, 
may nevertheless be ``similar'' to a quiet Sun. 

At the limb, the emission in \hedi\ shows a feature rising in a similar 
fashion as in \ion{Ca}{2}~H, but it reaches greater heights (almost 2~Mm higher, 
see Figure~\ref{fig:intt}). However,  the synthetic \hedi\  
disappears from the limb view in a different fashion than \ion{Ca}{2}~H. The 
latter disappears in a few seconds because the plasma is heated out of the 
passband. The synthetic \hedi\ at the limb does not vanish as 
fast as the synthetic \ion{Ca}{2}~H.

\begin{figure*}
\includegraphics[width=0.95\textwidth]{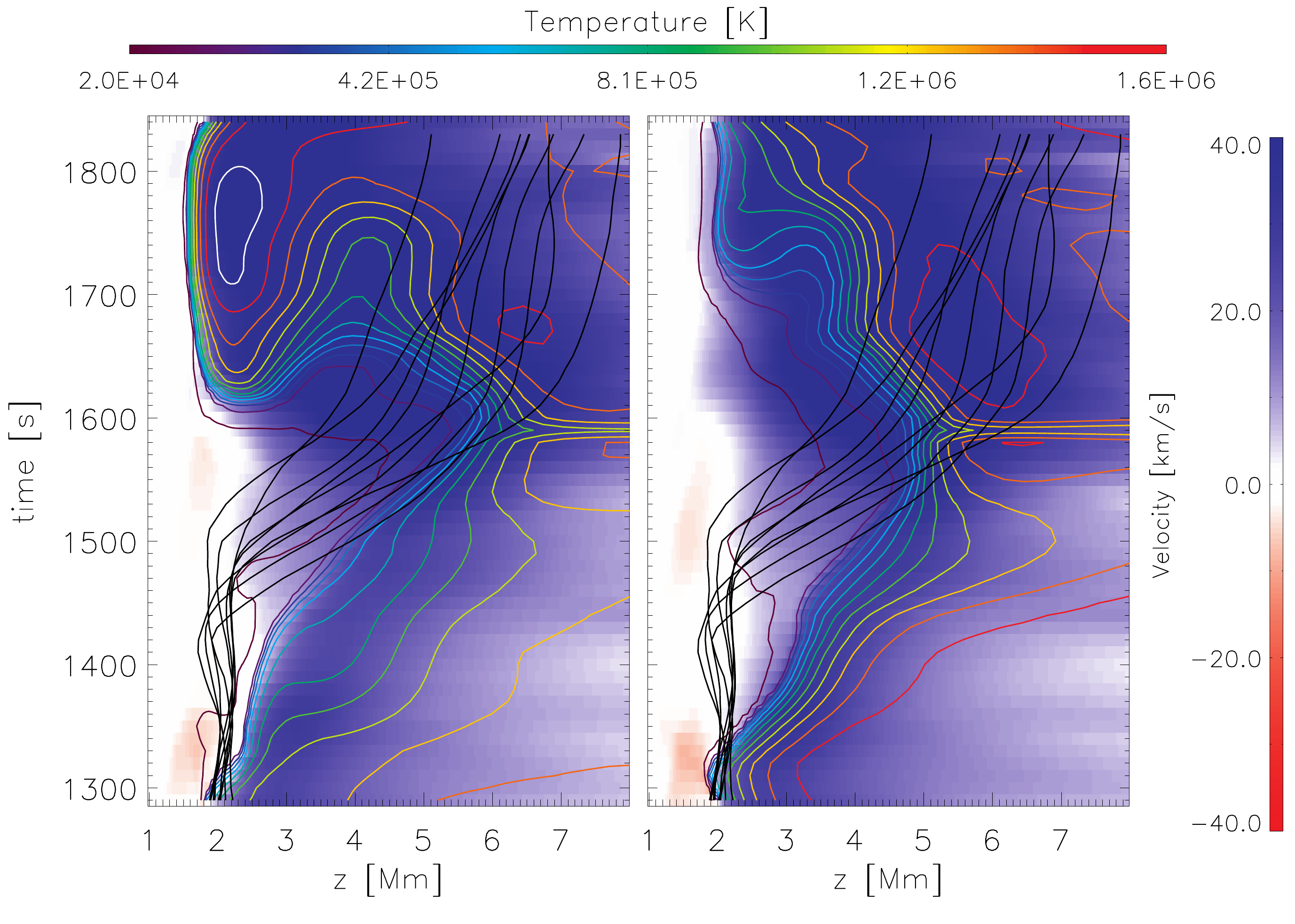}
\caption{\label{fig:tguz}  Vertical velocity (red-white-blue background color scheme) and temperature
	(color contours) are shown along a vertical axis centered in $y=6$~Mm and 
	$x=7.02$~Mm (left panel) and $x=6.83$~Mm (right panel). The projection of 
	particle trajectories are shown with black lines.}
\end{figure*}

The synthetic \hedi\  shows that the upper part of the spicule penetrates into 
the corona and at the same time fades until it disappears. While the 
typical observed behavior is for \hedi\ spicules to follow a parabolic profile 
with time \citep{De-Pontieu:2011lr}, a few examples in Figure~S5 in 
\citet{De-Pontieu:2011lr} actually show similar behavior to our simulation: e.g., at 
height $5$~Mm and $t=2200$s, $t = 2800$s, and $t = 3800$s. The reason for this 
evolution in our simulation is because the upper part of the spicule is gradually heated 
from above by Joule heating and conduction (see left panel of Figure~\ref{fig:tguz} 
and right panel of Figure~\ref{fig:caxz}). 

The observed parabolic height time profile of \hedi\ type II spicules has typically been interpreted as a consequence of plasma moving up and down, similar to the cooler dynamic fibrils and type~I spicules 
\citep{Hansteen+DePontieu2006,De-Pontieu:2007cr,Heggland:2007jt}. 
Our simulation suggests another possible scenario where a combination of mass 
flows, thermal evolution and narrow imaging passbands produce apparent up and 
down motion. In the right panel of Figure~\ref{fig:tguz} the temperature contour 
around the formation temperature of
\hedi\ (violet contours) appears to follow some sort of parabolic profile. When 
the spicule reaches its greatest height, the spicule starts to fade, which then 
causes the parabolic shape for transition region temperatures seen in space-time 
plots of intensity similar to the rapid fading of \Halpha\ or \ion{Ca}{2}~H described 
above. The heating for transition region temperatures is concentrated above the 
spicule and is not as rapid as in the location of the spicule shown in the left panel in 
Figure~\ref{fig:tguz}. The left and right panels are for two different locations in the spicule. 

For the other EUV lines, the emission follows the field lines forming the small loop as 
shown in Figure~\ref{fig:int}. The length of the synthetic spicule is a bit shorter 
compared with the real observations \citep{De-Pontieu:2011lr}, probably because:  
\begin{itemize}
\item The candidate presented here is not violent enough. 
\item \hedi\ ionization may be time-dependent.
\item Note that as a result of the small box size, 1) the magnetic field lines are 
confined in the small domain, 2) we do not have the same background 
contribution as in the observations, i.e., we lack the large LOS
integration at the limb that we find on the Sun.
\end{itemize}

\subsubsection{Line-width, Doppler shifts and asymmetries} 

\subsubsubsection{On the disk}
\label{sec:rbdisk}

\begin{figure*}
\includegraphics[width=0.99\textwidth]{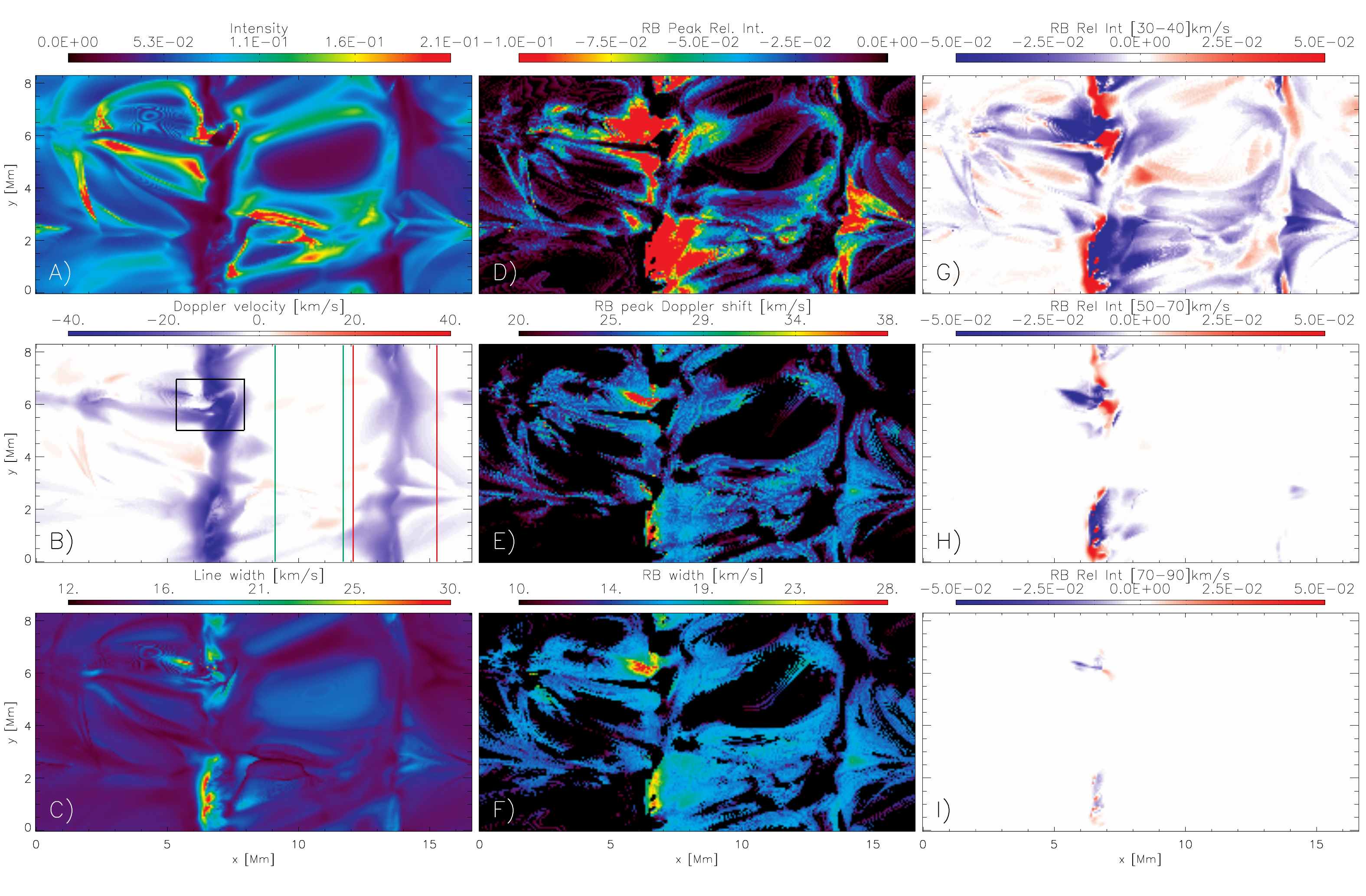}
\caption{\label{fig:rbmaps1} Spatial distributions of the parameters derived from 
	the single-Gaussian fit and RB asymmetry analyses for \sis\  at $t=1610$s. 
	The figure shows maps of the total intensity (panel A), Doppler velocity (panel B), 
	line-width (panel C), peak RB asymmetry (panel D), Doppler shifts of the RB 
	peak asymmetry (panel E), width of the RB asymmetry (panel F), and RB 
	asymmetry between [30-50], [50,70], and [70-90]~km~s$^{-1}$ (panels G-I). 
	The spicule is located inside the black box in panel B, the area between the 
	red lines delimits the footpoint of the transition region without any spicule and the 
	area between the green lines delimits a loop region 
	where the magnetic field is rather horizontal.}
\end{figure*}

The ejected chromospheric material reaches velocities up to 40~km~s$^{-1}$, 
but at higher temperatures  the plasma reaches larger and a wider range of upflow 
velocities \citep[as shown in Figure~3 in][]{Martinez-Sykora:2011uq}. A good 
diagnostic for analyzing these velocities is combining the information of the 
Doppler shift, line width and RB asymmetry as shown in Figure~\ref{fig:rbmaps1} for 
the spectral line \sis . The spicule is located at $[x,y]\approx [6,6]$~Mm. It shows 
Doppler shifts up to $50$~km~s$^{-1}$ (Panel B). The line width is larger than 
in the surroundings of the spicule and the profile shows a strong blueward 
asymmetry. Both the broad line and blueward asymmetry are a direct 
consequence of the strong spatial variation of the LOS velocity of the plasma 
that emits in this spectral line. In particular, in the spicule the line width is approximately 
32~km~s$^{-1}$. In contrast, the line width in the surroundings is only 
$18$~km~s$^{-1}$ or less (Panel C). The spicule shows blueward asymmetries of roughly 
10\% (Panel D) of the intensity centered at 50~km~s$^{-1}$ (Panel E). However, this 
blueward asymmetry is observed all the way up to 90~km~s$^{-1}$ (see Panels G-I).

\begin{figure*}
\includegraphics[width=0.99\textwidth]{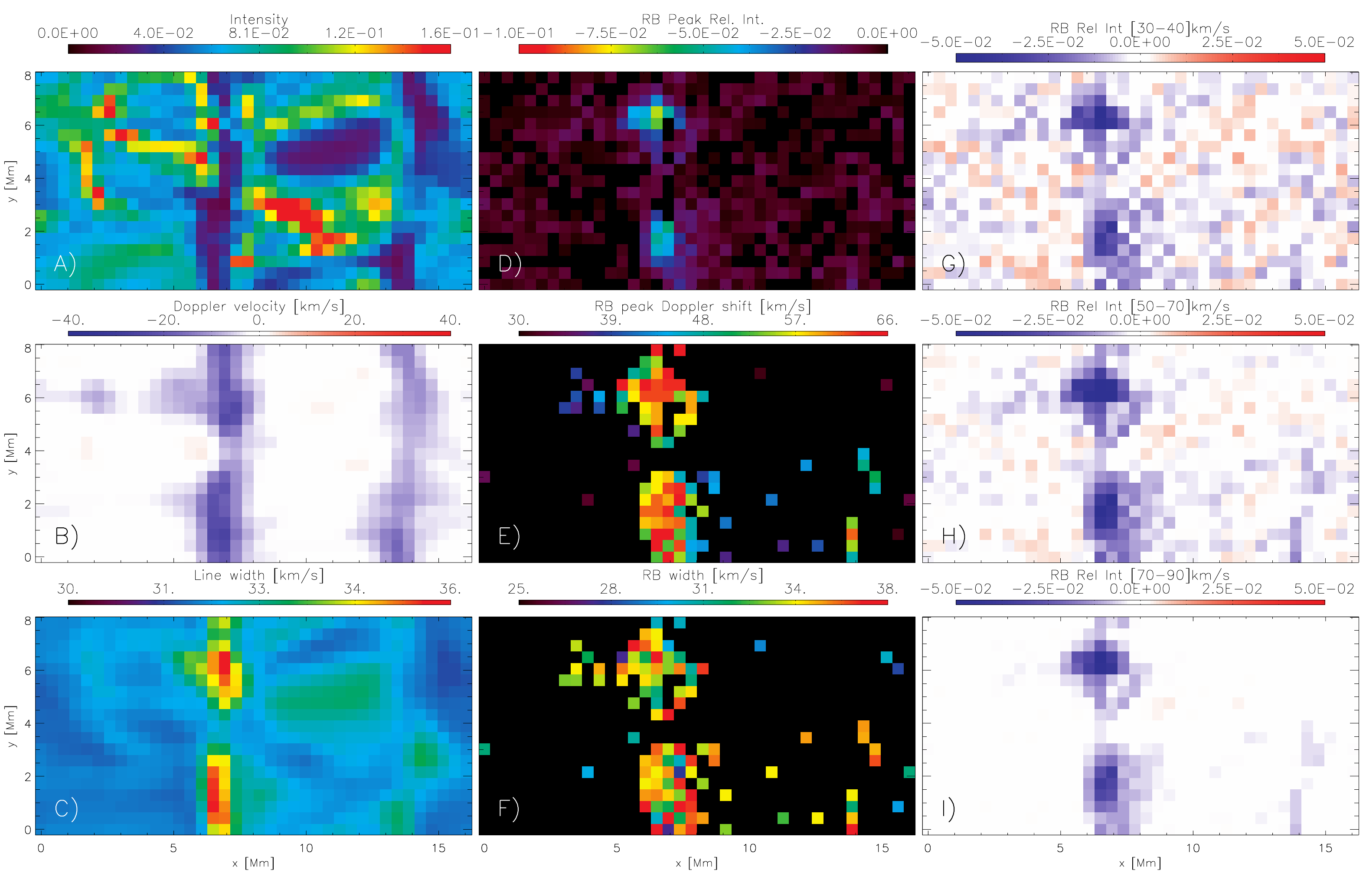}
\caption{\label{fig:rbmaps2}  Spatial distributions of the parameters derived from the 
	single-Gaussian fit and RB asymmetry analyses (as Figure~\ref{fig:rbmaps1}) 
	for \sis\ at $t=1610$s. These profiles are calculated taking into 
	account the Hinode spatial resolution and instrumental broadening of spectral lines.}
\end{figure*}

It is important to consider the instrumental effects on the spectral line,  
such as the spatial resolution and the instrumental broadening shown in 
Figure~\ref{fig:rbmaps2}. \citet{Martinez-Sykora:2011fj} described in detail the 
impact of these effects on the RB asymmetry diagnostics. As a result of the 
spatial resolution, the Doppler shifts are  significantly 
smaller ($\sim 25$~km~s$^{-1}$ for the spicule, panel B) because the profile is 
convolved with the pixel size. 
In general, the line-width is larger due to the instrumental broadening 
(panel C). However, at the location of the spicule, the line width is 
even larger. In addition, the RB asymmetry profile is shifted to higher  
velocities (panel E and panels G-I), and becomes wider (panel F). 
As result of the spatial averaging, the profiles include contributions from 
cells in the domain which leads to a wider distribution of velocity and 
temperature. In the spicule this leads to strong blueward asymmetries 
(see below). 

\begin{figure*}
\includegraphics[width=0.99\textwidth]{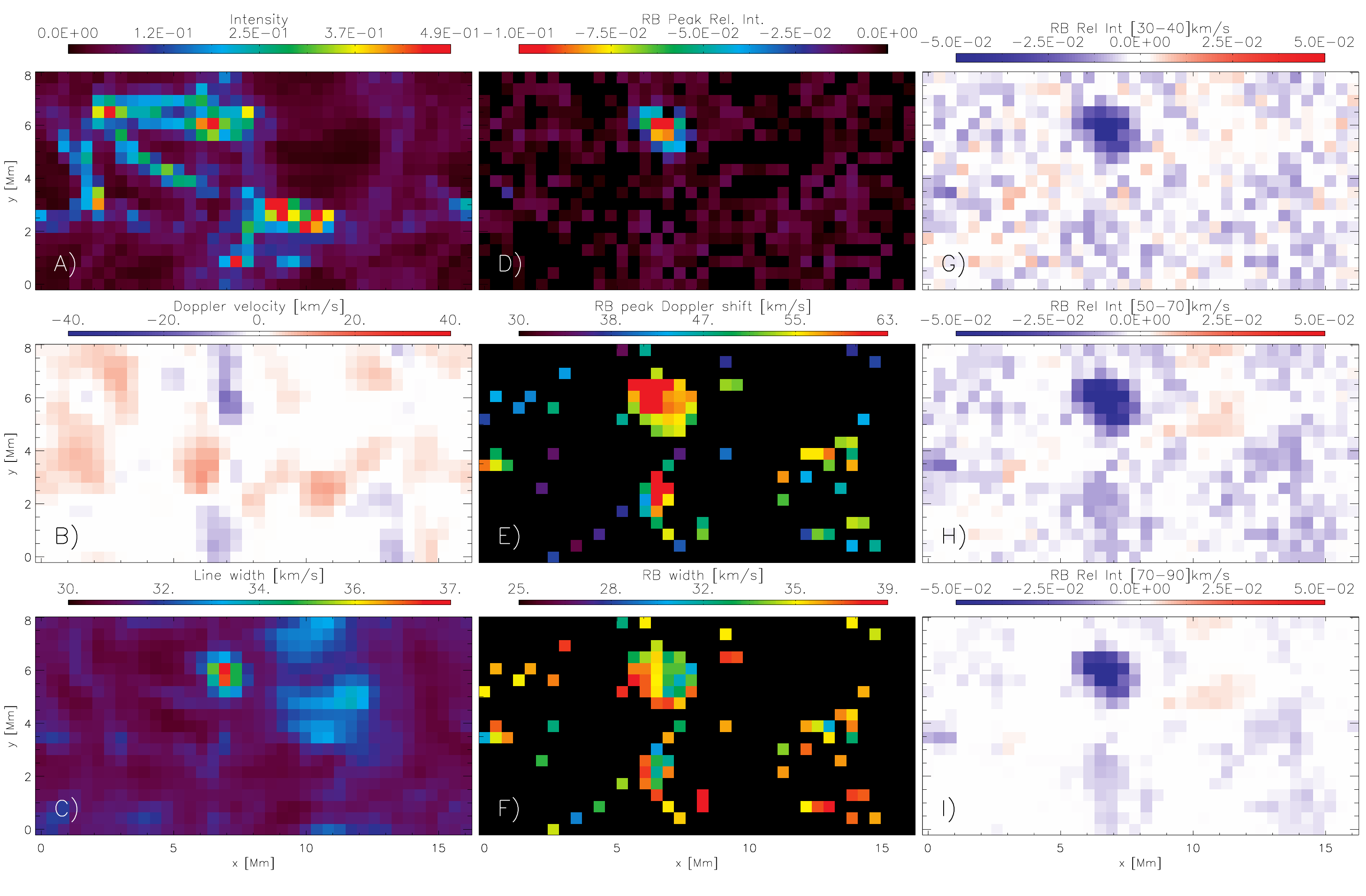}
\caption{\label{fig:n5rbmaps}  Spatial distributions of the parameters derived from the 
	single-Gaussian fit and RB asymmetry analyses (as Figure~\ref{fig:rbmaps1}) 
	for \nc\ at $t=1610$s. These profiles are calculated taking into account Hinode 
	spatial resolution and instrumental broadening of spectral lines.}
\end{figure*}

\begin{figure*}
\includegraphics[width=0.99\textwidth]{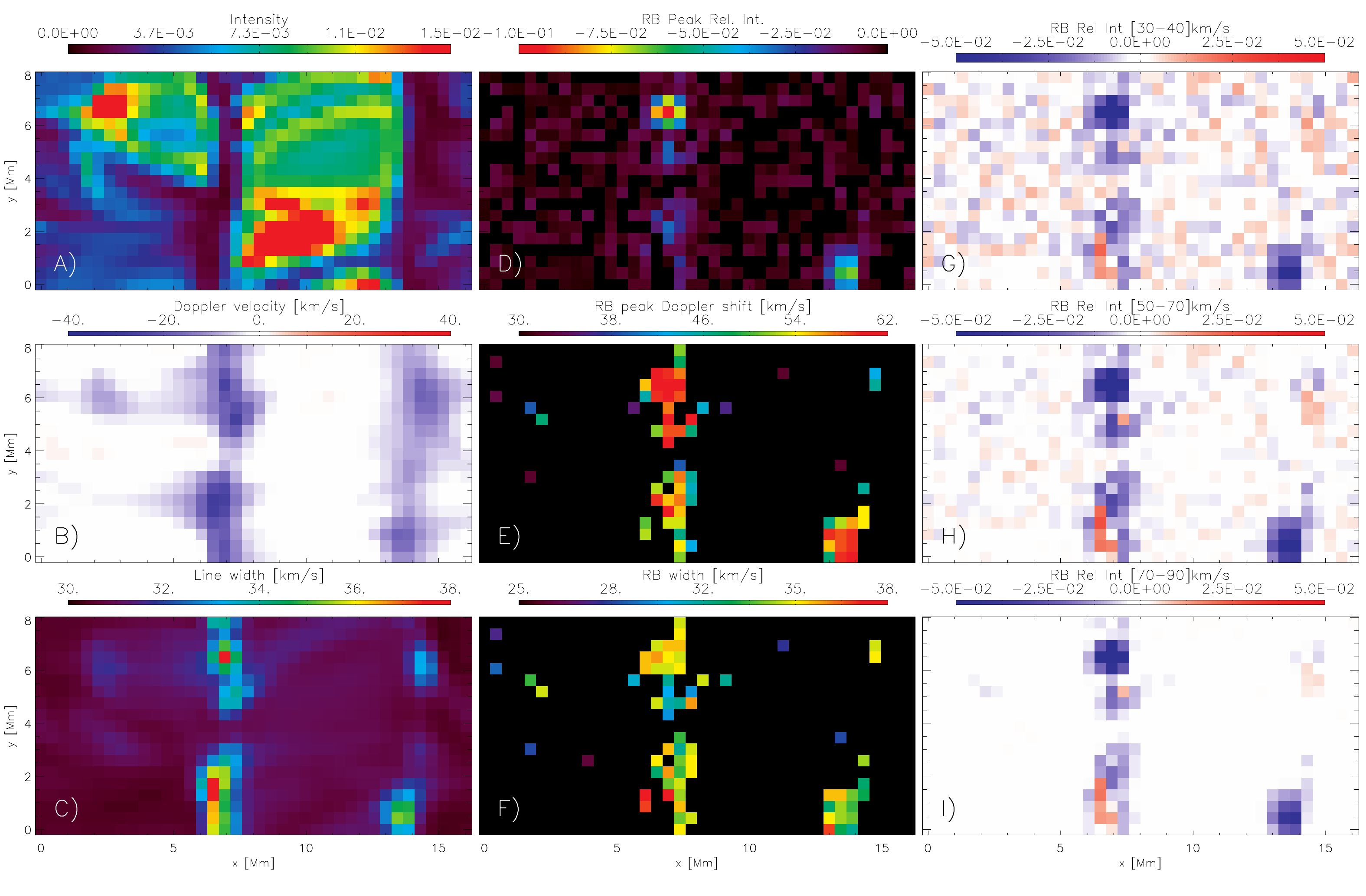}
\caption{\label{fig:fe12rbmaps}  Spatial distributions of the parameters derived from the 
	single-Gaussian fit and RB asymmetry analyses (as Figure~\ref{fig:rbmaps1}) for 
	\fet\ at $t=1610$s. These profiles are calculated taking into account Hinode 
	spatial resolution and instrumental broadening of spectral lines.}
\end{figure*}

In a similar manner as for \sis , transition region lines (see Figure~\ref{fig:n5rbmaps} 
for \nc ) and coronal lines (see Figure~\ref{fig:fe12rbmaps} for \fet ) show 
strong Doppler shifts, large line widths and blueward asymmetries of the order 
of 5\% located at $\sim50$~km~s$^{-1}$ at the spicule location. In the area of the 
spicule, the Doppler shift and asymmetry is a bit larger in \sis\ than in the 
other two lines. This is because of the thermo-dynamic properties of the spicule 
at that instant, which changes in time as described below. In addition, \sis\ is 
emitted by a Lithium-like ion. The atomic physics of the Lithium-like ions 
cause emission over a larger range of temperatures than for non-Lithium 
like ions \citep[see below and in][]{Martinez-Sykora:2011fj}. 

\begin{figure}
\includegraphics[width=0.45\textwidth]{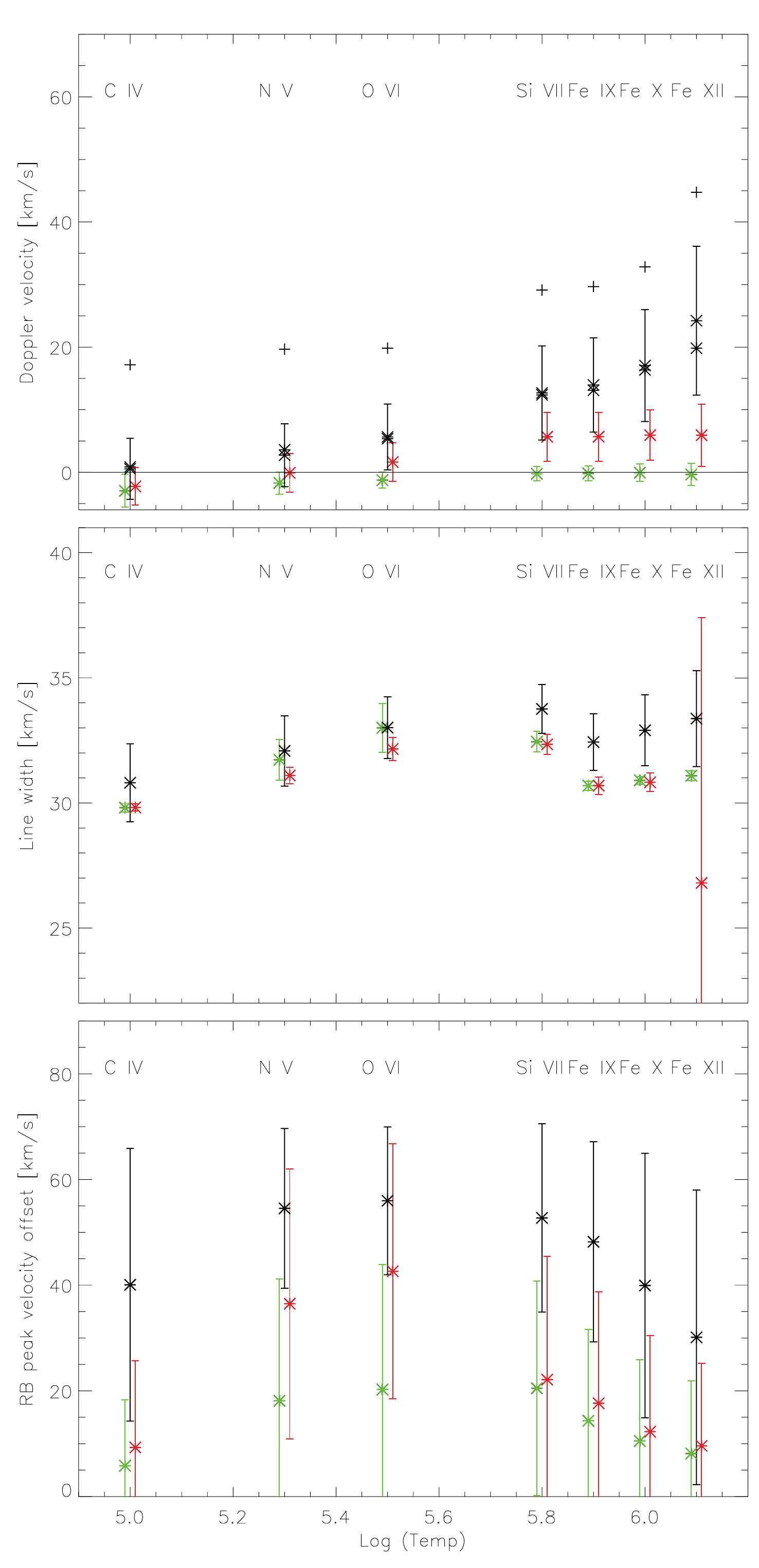}
\caption{\label{fig:dopp} Mean (asterisks) and standard deviation (error bars) of the 
	Doppler shift (top panel), line width (middle panel) and the Doppler velocity of 
	the RB peak asymmetry (bottom panel) at time $t=1600$s for different lines shown 
	as a function of the maximum formation temperature of each ion. The profiles have 
	been degraded to the spatial resolution of Hinode and instrumental broadening 
	of the spectral line profile. These values are calculated within three different 
	regions: the loop footpoints (the region delimited with the red square shown in 
	panel B in Figure~\ref{fig:rbmaps1}), the loop regions (the region delimited with 
	the green square in panel B in  Figure~\ref{fig:rbmaps1}),the spicule (the region 
	delimited with the black square) is shown in red, green, and black, respectively. 
	The $+$ symbol in the top panel refers to the maximum Doppler shift. The Doppler 
	velocity of the RB peak asymmetry is shown 
	only when the RB asymmetry is more than 0.5\% of the total local intensity.}
\end{figure}

At the footpoints of the region without type~II spicules (red square in panel B in 
Figure~\ref{fig:rbmaps1}), the Doppler velocity increases with the formation temperature 
of the various ions (red in the top panel of Figure~\ref{fig:dopp}). However, in the 
loop regions (the region delimited within the red square in panel B in Figure~\ref{fig:rbmaps1}) 
the Doppler velocity does not increase with the formation temperature of the ion(s) mainly 
because the plasma is confined to the magnetic field and it is mostly horizontal (green in 
the top panel of Figure~\ref{fig:dopp}). In the spicule, the Doppler velocities of all spectral 
lines (shown black in the top panel of Figure~\ref{fig:dopp}) are higher 
than the Doppler velocities in the other selected regions. The Doppler velocity seems 
rather constant with formation temperature of the ions for transition region lines 
(\cc , \nc\ and \os ), but for lines with higher formation 
temperatures the Doppler velocity increases significantly. 

Despite this significant rise of Doppler shifts with temperature, the actual 
vertical velocity in the spicule for the temperatures covered, does not increase
significantly with temperature. At coronal temperatures the spicular plasma at $t=1600$s 
shows only a small increase of the actual upflow velocity (as opposed to the 
Doppler shift, which is a property of the emergent line profile) with temperature 
as shown in Figure~3 in \citet{Martinez-Sykora:2011uq}. The modest increase in 
vertical velocity with temperature is considerably lower than what is expected from 
1D models \citep{Judge:2012uq}. There are several reasons for this: our simulated 
spicule arises as the result of Lorentz force driven acceleration that occurs along 
a large height range and acts differently for various locations within the jet. This complex 
and dynamic 3D structure defies simplification to the seemingly intuitive 1D interpretation. 
In addition, the simulated spicule expands with height which also acts to decrease the 
rise of the vertical velocity with temperature (panel B in Figure~\ref{fig:joucond}).

\begin{figure*}
\includegraphics[width=0.99\textwidth]{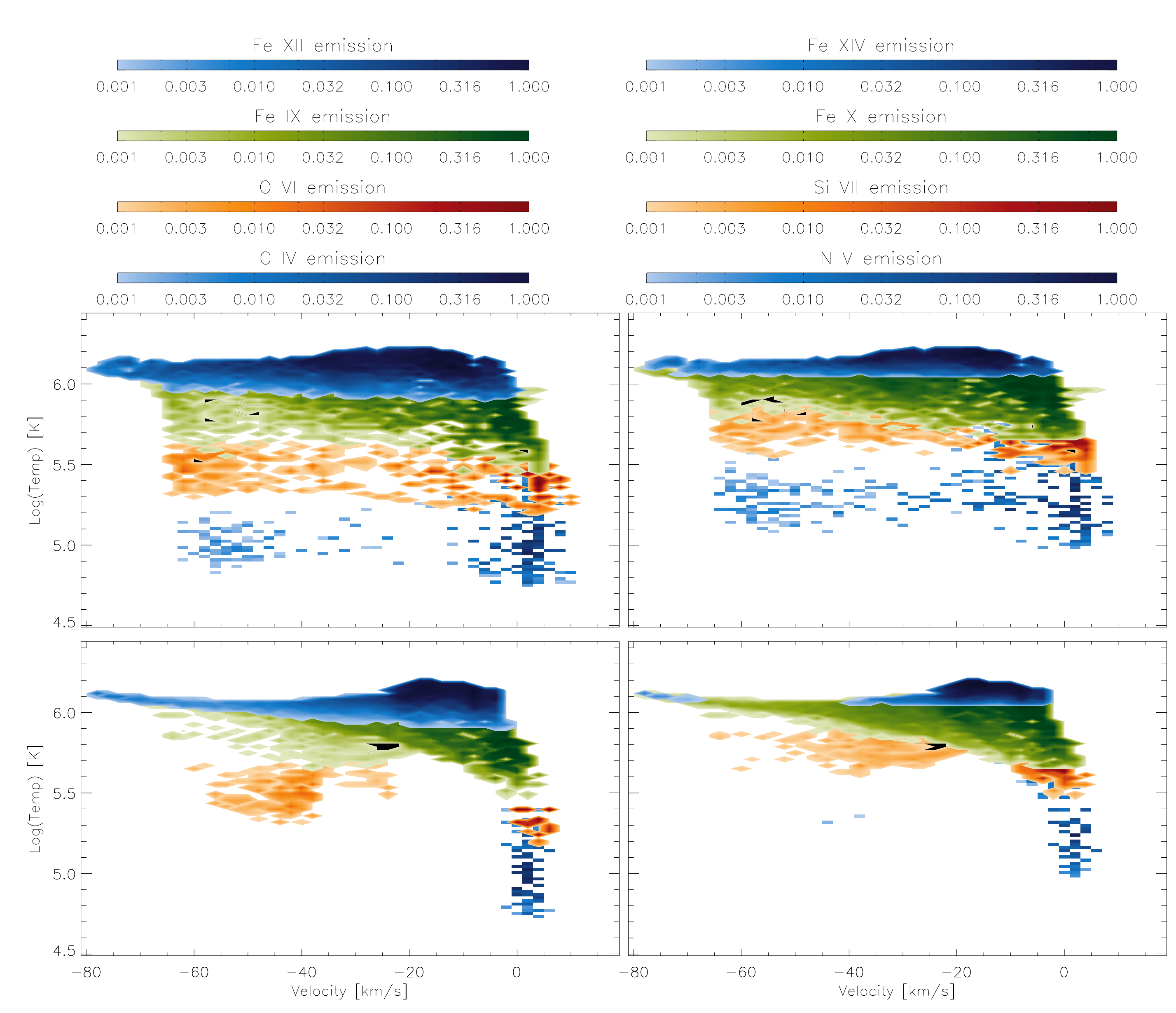}
\caption{\label{fig:tgemuz} Emission as a function of temperature and vertical 
	velocity in the region of plasma confined within one Hinode pixel 
	centered at $x,y=[5,6]$~Mm at $t=1600$s (top panels) and at $t=1680$s 
	(bottom panels). It is shown for different lines which follow different color 
	schemes displayed on top. \fed\ and \cc\ in the left panels and \fec\ 
	and \nc\ in the right panels share the same color scheme, but the emission 
	coming from \cc\ and \nc\ is centered around $Log(T)\sim5.0$, 
	and \fed\ and \fec\ are centered above $Log(T)\sim6.1$.}
\end{figure*}

The reason for the discrepancy between the actual vertical velocities and the Doppler 
shifts can be appreciated in the top panels of Figure~\ref{fig:tgemuz}. This figure 
shows the emission as a function of temperature and vertical velocity for the 
various lines. For the transition region lines (\cc , \nc, \os), 
the emission is concentrated in two velocity regions, one around 
$0$~km~s$^{-1}$ and the other around $60$~km~s$^{-1}$. The Doppler 
velocity of these profiles, using a single gaussian approximation, depends 
on the upward velocity and the contribution of the main emission and also 
the location and contribution of the second emission region. These two 
concentrations of emission are coming from the surrounding of the spicule 
(0~km~s$^{-1}$) and the upper part of the spicule (60~km~s$^{-1}$). In 
contrast, hotter (coronal) lines have a single region of emission spread over 
a large velocity range and the integrated emissivity decreases with increasing 
vertical velocity. The three main contributions to the coronal line emission are
the spicule, the hot loop associated with the spicule, and the hot corona overlying 
the simulated domain. Because of the large scale height of coronal lines 
we thus sample a wide range of velocities and emissivities. 
Figure~\ref{fig:tgemuz} clearly illustrates that using only the Doppler shift information 
of the line does not necessarily provide accurate information about the real 
velocity of the plasma at the different formation temperatures. In fact, using
only the Doppler shift can lead to the wrong conclusion: the Doppler 
velocity increases significantly for ions with the formation temperature but the 
vertical velocity of the plasma does not increase as much with temperature 
(Figure~\ref{fig:tgemuz}). Therefore, it is crucial to have a complete study of the 
line profile and include also the RB asymmetries (bottom panel in Figure~\ref{fig:dopp}). 
The coronal lines show stronger Doppler velocities because their emission 
does not have as much contribution from low velocities as the transition 
region lines and the strongest emission is located at higher velocities 
\citep{Martinez-Sykora:2011fj}. As a result of this, the blueward RB asymmetries 
of the coronal lines are shifted to lower velocities 
compared to the transition region lines. We note that this conclusion is valid
for the current simulation, which does not have much overlying coronal 
emission. On the Sun, the impact of the LOS superposition on 
Doppler shifts and line asymmetries will critically depend 
on the overlying background emission. 

Inside the spicule, the line width of the spectral lines is larger for the various 
lines shown in the middle panel of Figure~\ref{fig:dopp} than in the other selected 
regions. The line width also increases with increasing formation temperature up 
to  $1.2\,10^{6}$~K. This linear increase is becomes smaller around at $\log(T) = 5.8$ because
the hotter lines are emitted by non-Lithium like ions which have a different 
temperature dependence of the contribution function ($G(T,n_{e})$). 
For instance, Lithium-like ions (\cc , \nc , \os , and \sis ) show a significant high-temperature 
tail of the $G(T,n_{e})$ compared to non-Lithium like ions, such as 
\ion{Fe}{9}-XII \citep[see Figure~13 in][]{Martinez-Sykora:2011fj}. As a result of this, 
Lithium-like ions show contributions coming from plasma with a wider range of 
temperatures than non-Lithium like ions; and the line width will be larger for the 
Lithium-like ions as a result of the LOS integration. Such behavior can be seen  
also in other regions such as the loop region or the transition region footpoints. 

\begin{figure*}
\includegraphics[width=0.98\textwidth]{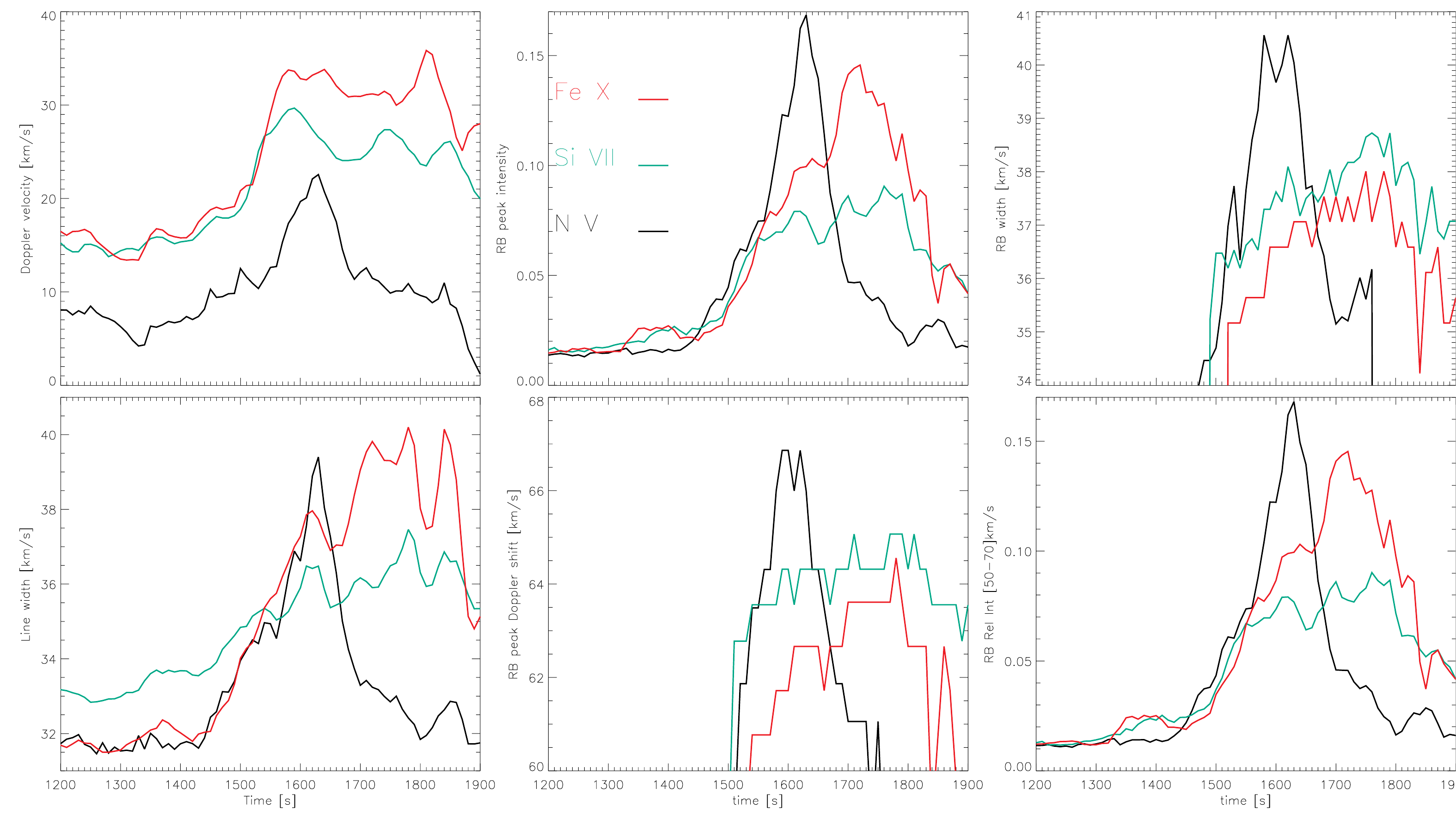}
\caption{\label{fig:evol} Time evolution of the Doppler velocity, line width, RB 
	asymmetry intensity, the Doppler shift of the RB asymmetry peak, Doppler 
	width of the  RB asymmetry profile and the RB asymmetry intensity around 
	50-70~km~s$^{-1}$ from top to bottom and left to right respectively. A  
	representative transition region spectral line (\nc\ in black), a representative 
	coronal line (\fet\ in red) and an intermediate line between these two (\sis\ in 
	green) are shown. The profiles have been degraded to the spatial resolution 
	and instrumental broadening of Hinode lines. The line parameters have been
	calculated by averaging over a square of 3x3 pixels where the spicule 
	is located.}
\end{figure*}

The time evolution of the properties of a representative transition region spectral line 
(\nc\ in black), a representative coronal line (\fet\ in red) and an intermediate line 
(\sis\ in green) are shown in Figure~\ref{fig:evol}. In order to calculate the velocity 
(bottom-middle panel) and width (top-right panel) of the RB 
asymmetry profile we limit this to RB asymmetries larger than 0.5\% of the peak intensity. 
Before the spicule emerges all the lines show blue Doppler shift (positive top-left panel) 
which is larger for ions with higher formation temperature. At the time the 
spicule appears (t=1470s), the Doppler shift increases with time for all lines, 
and the increase is largest for coronal lines. However, it is interesting that the 
duration of high Doppler shift is shorter for transition region lines (\cc, 
\nc , and \os ), than for the coronal lines (\fen, \fet , and \fec ). Similar  behavior 
is observed also in the other properties of the spectral lines such as the line 
width (bottom-left panel), the RB intensity (middle top), the velocity of the RB 
peak asymmetry (bottom-middle panel), the width of the RB asymmetry profile 
(top-right panel) and the RB asymmetry around 50-70~km~s$^{-1}$ (bottom-right panel). All these 
properties increase at the same time as the spicule evolves but this increase 
is shorter-lived for the transition region lines than for the coronal lines. 
This suggests that trying to estimate the lifetime of the coronal
counterparts of spicules based on TR diagnostics might lead to
misleading results. In particular, it suggest that 
the results of \cite{Klimchuk:2012fk}, who cast doubt on the impact of
spicules on the corona, likely significantly underestimate the real
lifetime of the coronal counterpart of spicules. 

During the early phase of the spicule most of the spicular plasma is at 
chromospheric and transition region temperatures and has a wide range
of velocities because of the complex acceleration mechanism. This leads
to strong Doppler shifts, line widths and asymmetries. The coronal emission 
``riding'' on top of the spicule will also show a similar increase of the spectral 
line parameters. Later in time, the spicule is heated and the transition region 
lines show only the footpoints of the spicule, whereas the coronal emission 
occurs throughout the spicule. As result of this, the transition region lines 
show the emission of a limited volume (height) where the velocities 
are all very similar and small as shown in the bottom panels of 
Figure~\ref{fig:tgemuz}. Therefore, the LOS integration is limited to a thinner 
region and as a consequence, the Doppler velocity, the line width, and the RB 
asymmetries no longer show an increase in time for transition region lines. In contrast, the 
coronal lines {\em maintain} the RB asymmetries as a result of the dynamics 
of the spicule (upflows) and remain even after the chromospheric material 
is heated. Note that if time dependent ionization of the various ions is taken 
into account, the diagnostics may change 
\citep{Hansteen:1993kx,Bradshaw:2009uq,Judge:2012uq,kosovare:2012}.

It is also interesting that the transition region lines show the strongest 
asymmetries, and largest shift and width of the RB peak asymmetry. As 
mentioned above, for transition region lines, the emission comes from two  
regions, one in the footpoints of the spicule and the second one from the top 
of the spicule. As a result of this, the asymmetries are larger than in coronal lines. 
We want to highlight again the importance of studying the Doppler shifts 
together with the asymmetries of the spectral lines. Failure to do so can easily 
lead to incorrect conclusions. 

It is crucial to mention that in the synthetic observables we do not have very 
strong background emission because of the small size of the computational 
domain. The background emission impacts the properties of the profiles such 
as Doppler shifts, and asymmetries \citep{Martinez-Sykora:2011fj}.

\subsubsubsection{At the limb}

\begin{figure}
\includegraphics[width=0.49\textwidth]{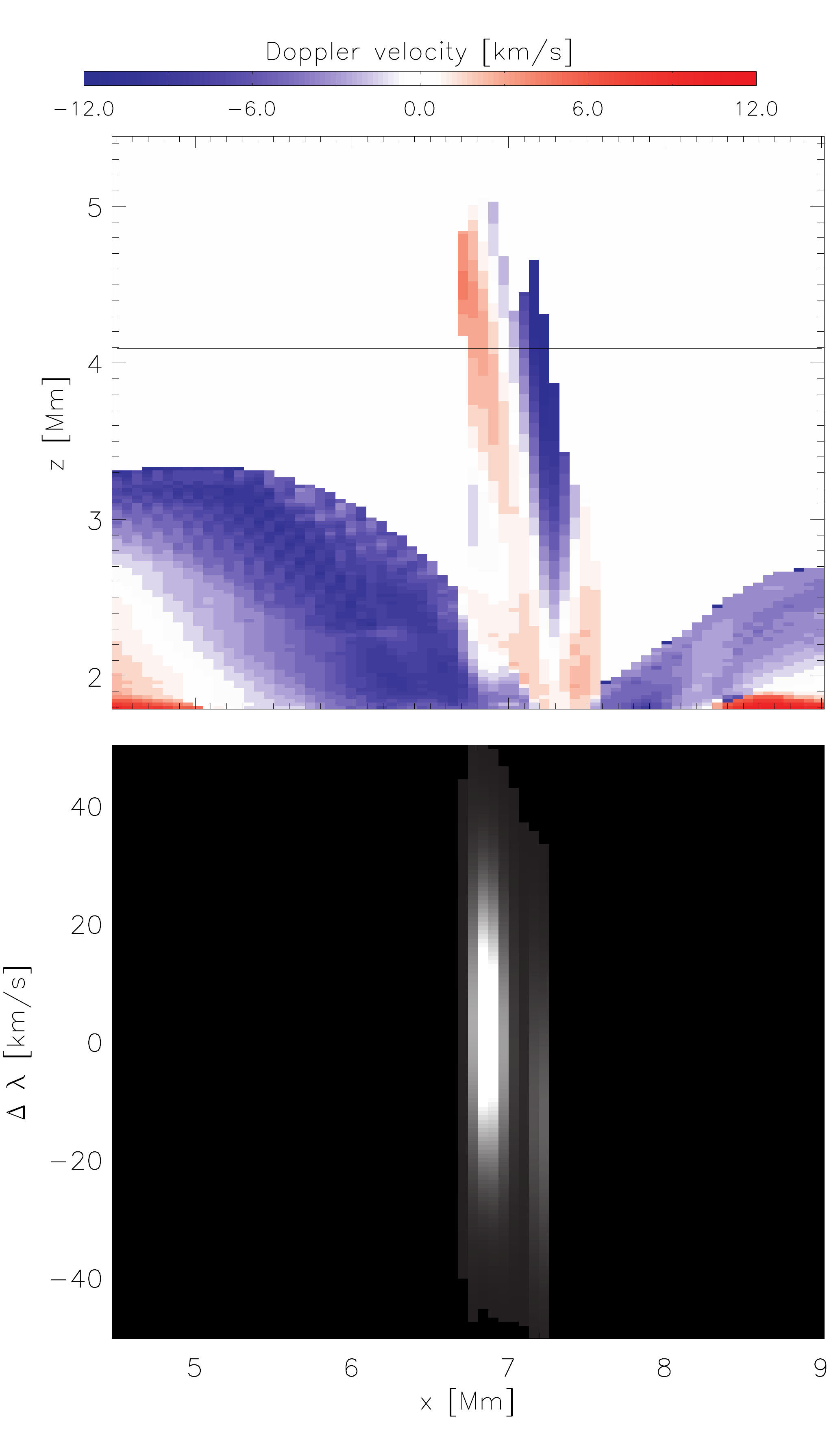}
\caption{\label{fig:hadopxz} Off-limb view of large torsional motion in the 
simulated spicule in \Halpha. 
Doppler shift map at the limb (i.e., the xz-view) is shown in the top panel and 
the spatial variation of the \Halpha\ profile at $z=4.1$~Mm is shown in the bottom panel.}
\end{figure}

At the limb, the Doppler velocity map of synthetic chromospheric lines such as \Halpha\  
shows red and blue shift at each side of the spicule (see Figure~\ref{fig:hadopxz}) as 
if the plasma is rotating at the same time that it is ejected into the corona.  The Doppler 
velocities observed at the limb are of the order on $\pm10$~km~s$^{-1}$. 
Such torsional motions have recently also been observed by 
\citep{De-Pontieu:2012bh} in \ion{Ca}{2}~H. The Doppler velocities are larger in 
the observations than the velocities observed in the synthetic profiles. 
However, the difference seems similar to the discrepancies between the 
various properties of the synthetic RBE (e.g., Doppler velocities in the disk view) in 
comparison with the observations. Once again, these results suggest similarities with 
the observations but also indicate that 
this simulated spicule is not as violent as the type~II spicules on the Sun. 

Another main difference between the observations and the synthetic profile is 
that in the observations \citep{De-Pontieu:2012bh} the spatial variation of the 
spectral profile (bottom panel of Figure~\ref{fig:hadopxz}) does not show 
two treads. Such threads are seen in the synthetic case. These threads appear as a 
consequence of the variation of density within the spicule. The multithreaded 
nature of the simulated spicule is however compatible with suggestions of 
such structuring in observations \citep{Suematsu:2008zr}. 

\clearpage

\begin{figure*}
	\includegraphics[width=0.99\textwidth]{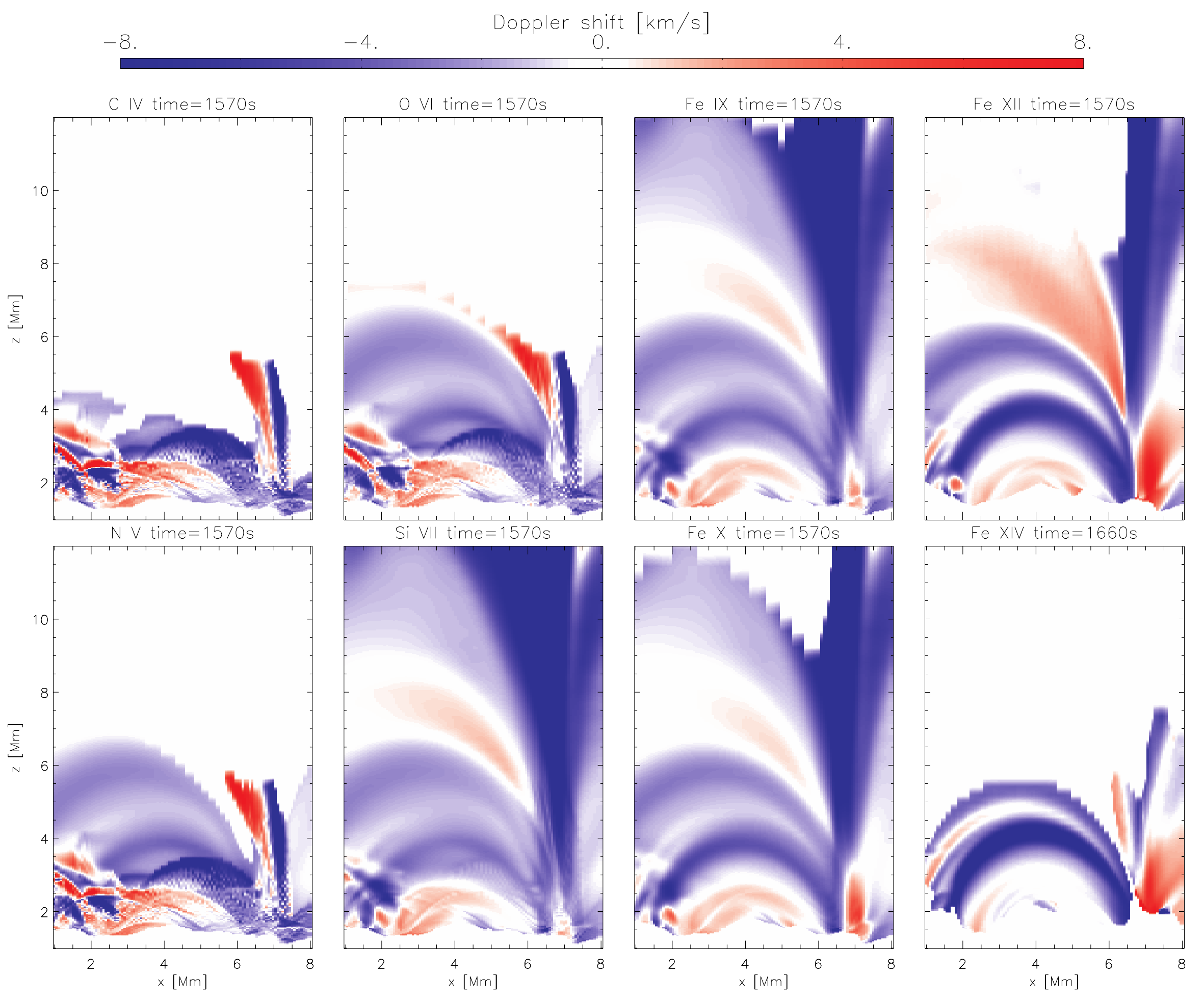}
	\caption{\label{fig:dopxz} Doppler shift maps at the limb (i.e., the xz-view) for different 
	lines. Note that the spicule shows in the left side a red-shift and in the right side a 
	blue-shift.}
\end{figure*}

This Doppler shift pattern at the limb is observed also in transition region and coronal 
lines as shown in Figure~\ref{fig:dopxz}. This effect can be seen in any 
LOS direction at the limb, i.e., when integrating along the x-axis instead of
the y-axis the red-blue Doppler shift structure is still visible. This Doppler shift is not 
uniform or constant in space as a function of the formation temperature of the various lines, 
and also the evolution seems to differ between the various lines. The transition 
region lines show such Doppler shift structure only when the chromospheric material is injected 
into the corona, before it is heated. The Doppler shifts in the transition region lines are larger 
than in the coronal lines but smaller than in the chromospheric lines. In addition, 
for \ion{Fe}{9}-XII this Doppler shift structure seems to last for a shorter time. For \fec, the 
red-blue Doppler shift appears later because the plasma needs to be heated to high 
temperatures. It also lasts longer than for the other cooler coronal lines. 

\begin{figure}
	\includegraphics[width=0.99\textwidth]{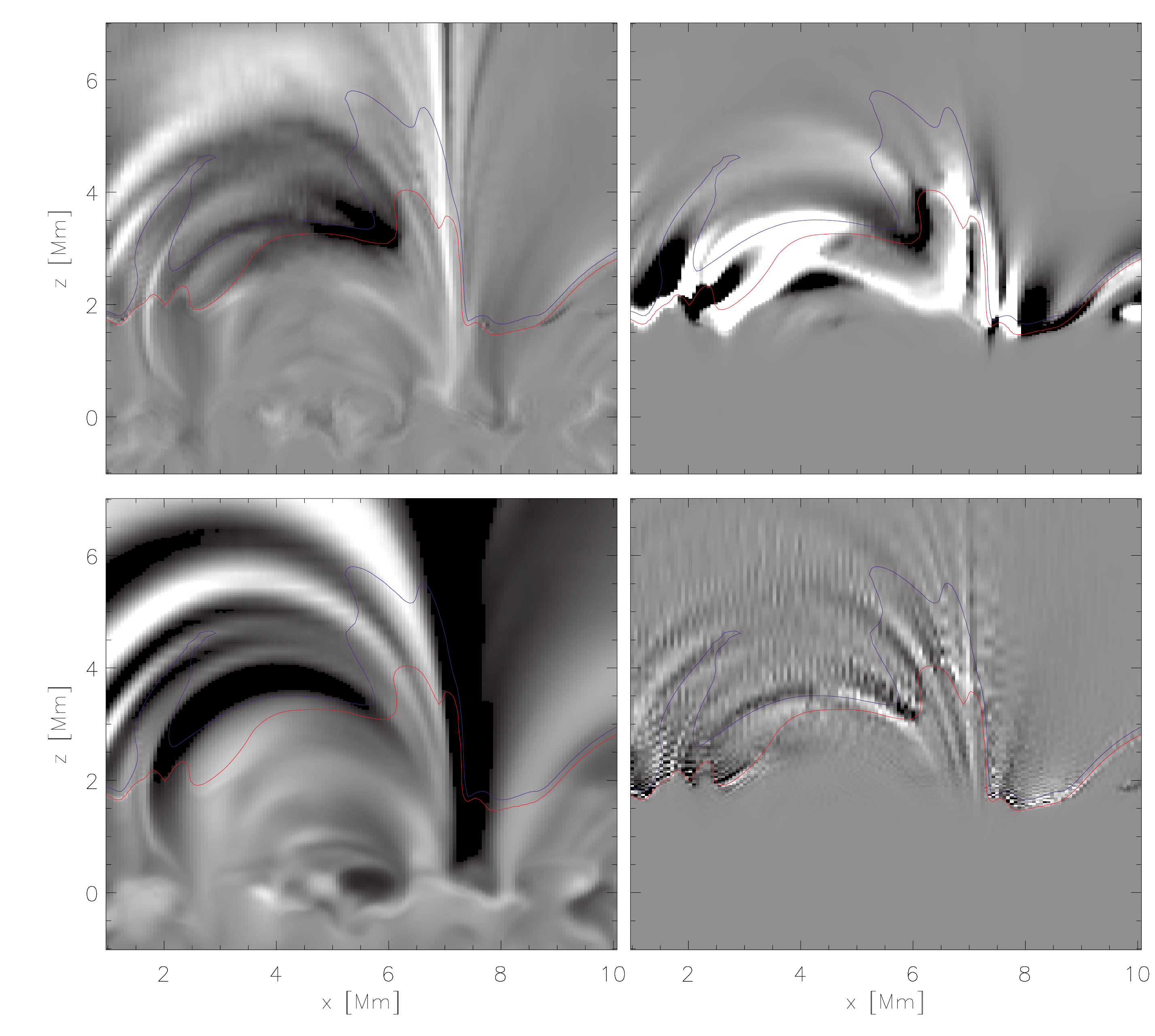}
	\caption{\label{fig:twist} xz cut of the vertical vorticity (top-left panel, [-7:7]~s$^{-1}$), 
	the velocity perpendicular to the plane ($u_{y}$, bottom-left panel, 
	[-5:5]~km~s$^{-1}$), the magneto baroclinic term 
	(top-right panel) and the magnetic tension (bottom-right panel) in an early stage of 
	the spicule ejection ($t=1490$s). The latter two plots show a range covering 
	[-2:2]~s$^{-2}$. The temperature contour at $2\,10^{5}$ and $5.5\,10^{5}$~K 
	are overlaid with red and blue colors respectively. }
\end{figure}

The Dopplershifts in the various limb views suggest the spicule is rotating. 
We confirm this by studying the vorticity in the simulations as shown in Figure~\ref{fig:init} 
with the velocity streamlines, but what causes it to rotate in the 
chromosphere and corona? One can derive a conservation equation of the 
vorticity by taking the curl of the MHD equations:

\begin{eqnarray}
&&\frac{D {\bf \omega}}{D t} =({\bf \omega} \cdot \nabla) {\bf u} -{\bf \omega} ( \nabla {\bf u}) +\frac{1}{\rho^{2}}\nabla \rho \times \nabla p  \nonumber \\
&&+\frac{1}{\rho^{2}}\nabla \rho \times \left[\nabla p_{mag} -  \frac{1}{4\pi}({\bf B} \cdot \nabla){\bf B}\right] \nonumber \\
&& +\frac{1}{4\pi \rho}\nabla \times ({\bf B} \cdot \nabla){\bf B} \label{eq:vort}  \\
\end{eqnarray}

\noindent where ${\bf \omega}= \nabla \times {\bf u}$, $p$, $\rho$, ${\bf B}$, 
and $p_{mag} = |B^{2}|/(4 \pi)$ are the vorticity, gas pressure, density, 
magnetic field, and the magnetic pressure. The terms in the right side of the 
equation are the tilting, stretching, baroclinic, magneto baroclinic terms, 
and magnetic tension \citep{Shelyag:2011uq,Steiner:2012fk}. 
Figure~\ref{fig:twist} shows in an early stage of the spicule ejection the 
vertical cut of the vorticity, the velocity perpendicular to the plane ($u_{y}$), 
and the two most important terms on the right side of the Equation~\ref{eq:vort} 
in the chromosphere and corona, i.e., the magneto baroclinic term and the 
magnetic tension. As a result of the compression at the footpoint of the spicule 
due to the magnetic field tension, the magneto baroclinic term generates a vorticity 
not only inside the chromosphere but also all the way up into the corona. This 
vorticity is generated in the chromosphere at the footpoint of the 
spicule, and it is the result of the compression. The magnetic baroclinic term is 
important because the density decreases along the y-axis and this density 
gradient is perpendicular to the gradient of magnetic pressure that is 
perpendicular to the magnetic field. 
In addition, the gradient of magnetic pressure perpendicular to the field 
lines is large along the x-axis. This is a consequence of the plasma being 
compressed: resulting in the generation of vertical vorticity. 

\subsection{Comparison with observations}~\label{sec:compobs}

\begin{figure*}
	\includegraphics[width=0.99\textwidth]{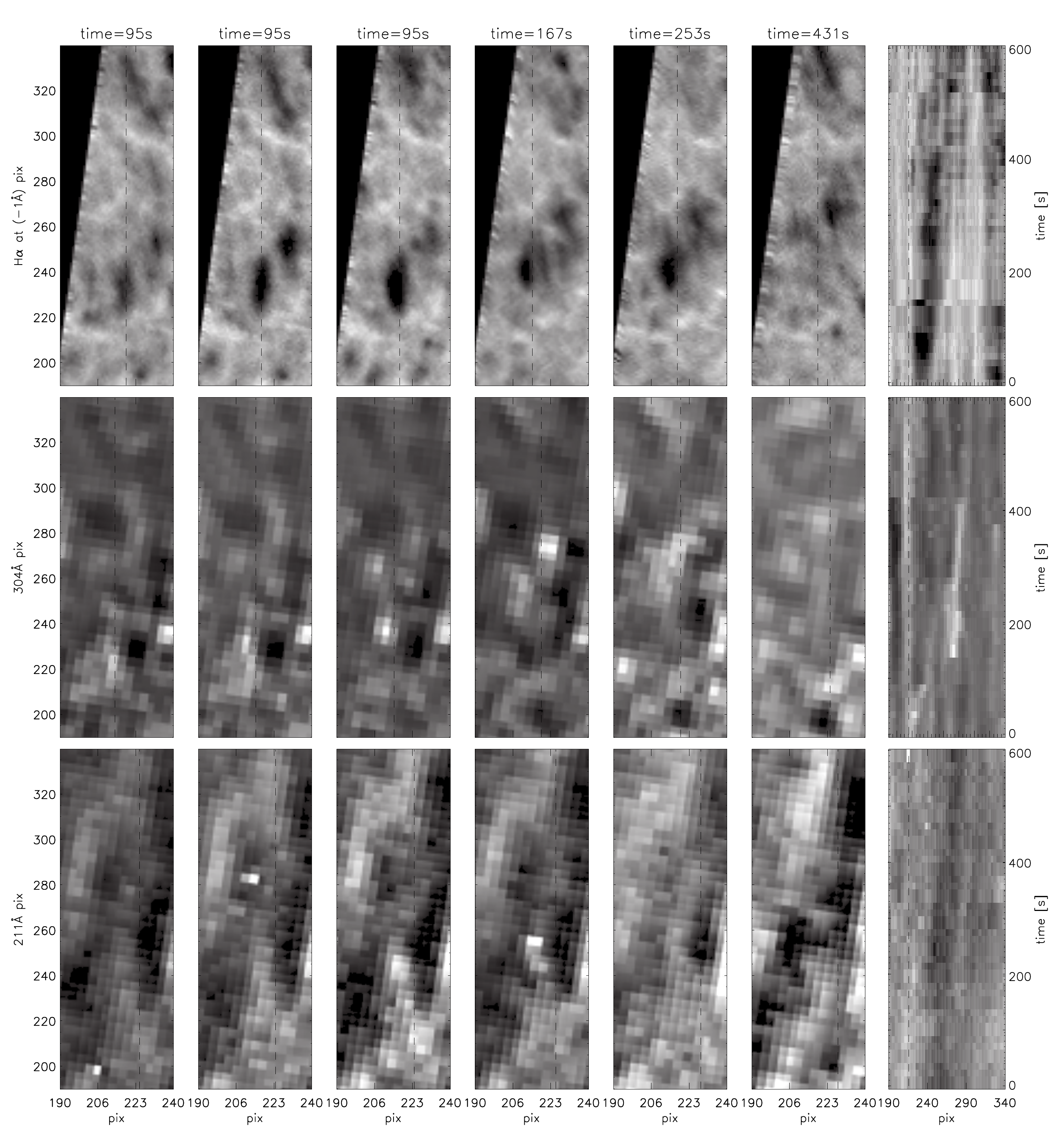}
	\caption{\label{fig:aiaobs} Temporal evolution of an observed 
	RBE visible as a dark feature in \Halpha\ (-868m\AA ) using Hinode, and associated 
	brightenings in 304\AA\ and 211\AA\ intensity using SDO/AIA  shown from 
	top to bottom in the left panels. Space time plots(right most column) along the axis of 
	the RBE (dashed lines in the left panels) show a brightening in the 
	transition region and coronal images along the spicule but also at the footpoint of 
	the spicule. The brightening at the footpoint in 304\AA\ and 211\AA\ lasts for a few
	hundred seconds. }
\end{figure*}

\citet{De-Pontieu:2011lr} used SDO/AIA observations to show the impact of  
type~II spicules into the corona using SDO/AIA. Here we use the same data set 
to see if we can recognize some similar features in the simulated spicule. 
Figure~\ref{fig:aiaobs} shows a time series of an RBE visible as a dark feature in 
\Halpha\ (-868m\AA ) (top row). Along the RBE the second and third rows of the 
figure show a brightening at transition region (304\AA\ channel) and coronal 
(211\AA\ channel) temperatures \citep{De-Pontieu:2011lr}. Note that the features 
are not overlapping exactly in space and time. This property agrees with the 
simulation. In the simulation, the emission from different lines shows different 
structures and do not overlap because the heating is not uniformly located 
and the various temperature structures in the spicule and surroundings are 
complex. As mentioned before, the fact that the brightenings do 
not share the exact location and instant implies that the 304\AA\ channel most likely  
shows transition region temperatures (\hed ) and the 211\AA\ channel shows 
coronal temperatures (\feci ). 

\begin{figure*}
	\includegraphics[width=0.99\textwidth]{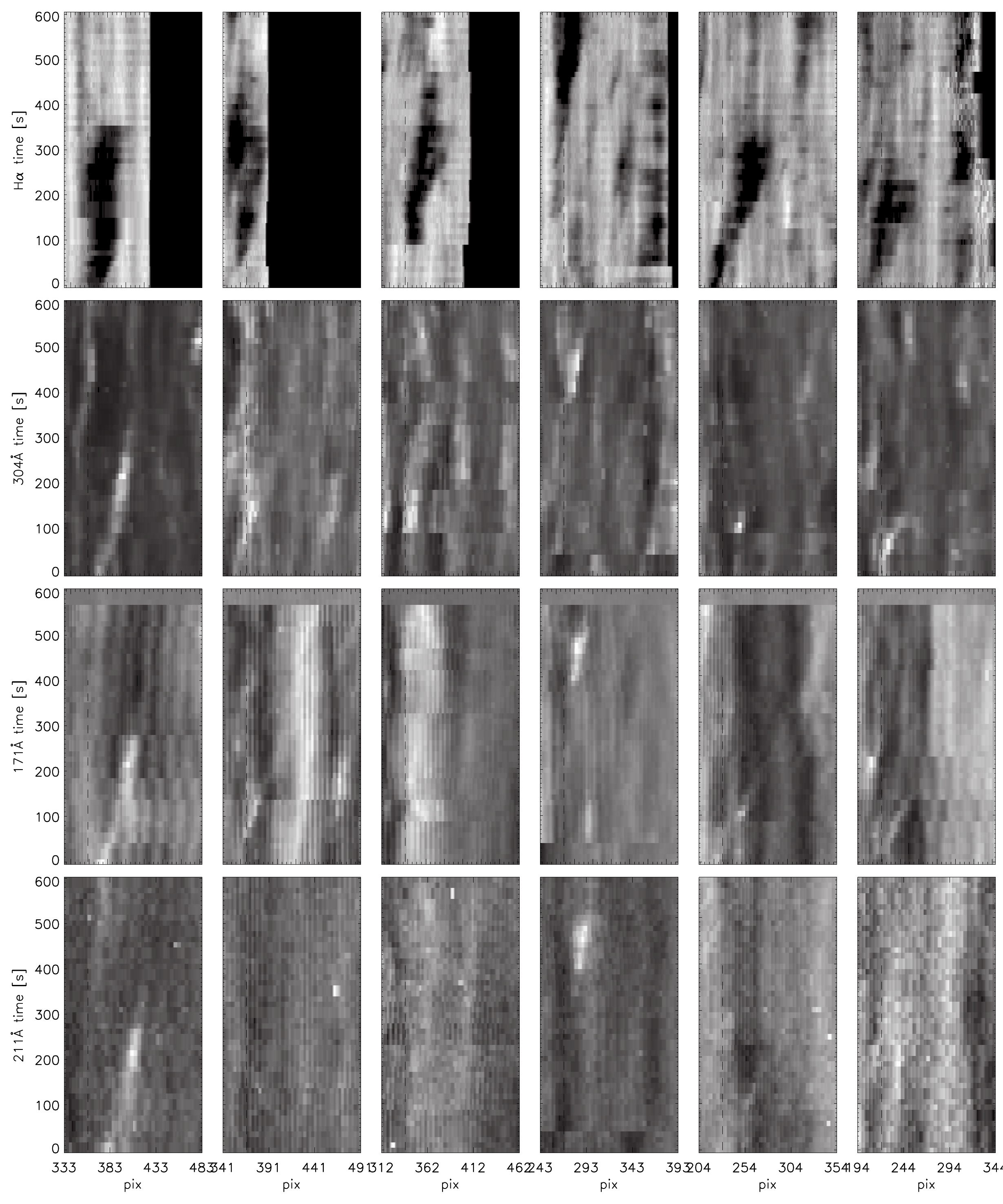}
	\caption{\label{fig:maiaobs} Space time plots along the axis of the RBE (top row)
	showing brightenings in transition region (second row, 304\AA ) and coronal 
	images (third and fourth row, 171\AA\ and 211\AA ) during the evolution of 
	several spicules. In some AIA channels such brightenings also occur at the 
	footpoints of the various spicules. The 
	brightenings at the footpoints last for a few hundred seconds.}
\end{figure*}

There are two major features that catch our attention in the time evolution of the 
brightening in the AIA channels shown in Figure~\ref{fig:aiaobs}. One of these is hardly 
seen in the modeled spicule. This brightening is following the spicule in the SDO/AIA channels 
and disappears almost as quickly as the spicule fades. This was observed by 
\citet{De-Pontieu:2011lr}. This brightening is small in the simulation
in comparison with the observations. This may be a consequence of the simulated 
spicule being not as violent as the observed spicules. The other feature seems to 
be more evident in the simulations than in the observations but still observed here in  
the various SDO/AIA channels. This is the brightening at the footpoint of the RBE 
which lasts for a few hundred seconds, and seems to increase in time. 
This may indicate that the heating mechanism 
persists longer than the lifetime of the RBE. Other examples of this type of 
observations are shown in Figure~\ref{fig:maiaobs}, which shows time-space 
plots where these two features in the time evolution can be seen in 
one or more of the SDO/AIA channels. Note  
that these data are pushed to the limit of their spatial and temporal 
resolution and signal to noise to enhance the brightening. We find that in the 
AIA observations it is more common to see an increase of brightening in the 
304\AA\ channel   than in the coronal emission. In contrast, in the simulation we 
see the opposite. This discrepancy could be explained by two issues. First, the 
modeled magnetic field configuration is different from that on the Sun. In the 
simulation the magnetic field is confined to a small domain and the magnetic 
field lines that produce this strong coronal emission close within the computational 
domain. Therefore, the heat and high temperatures are confined on these lines, 
so thermal conduction does not strongly dilute the heat input. This may lead to a 
modeled coronal emission that is brighter than on the Sun. Second, the formation of 
\hed\  is not well understood \citep{Feldman:2010lr}, which likely leads to discrepancies 
between our calculations and the observations. 

\begin{figure*}
	\includegraphics[width=0.99\textwidth]{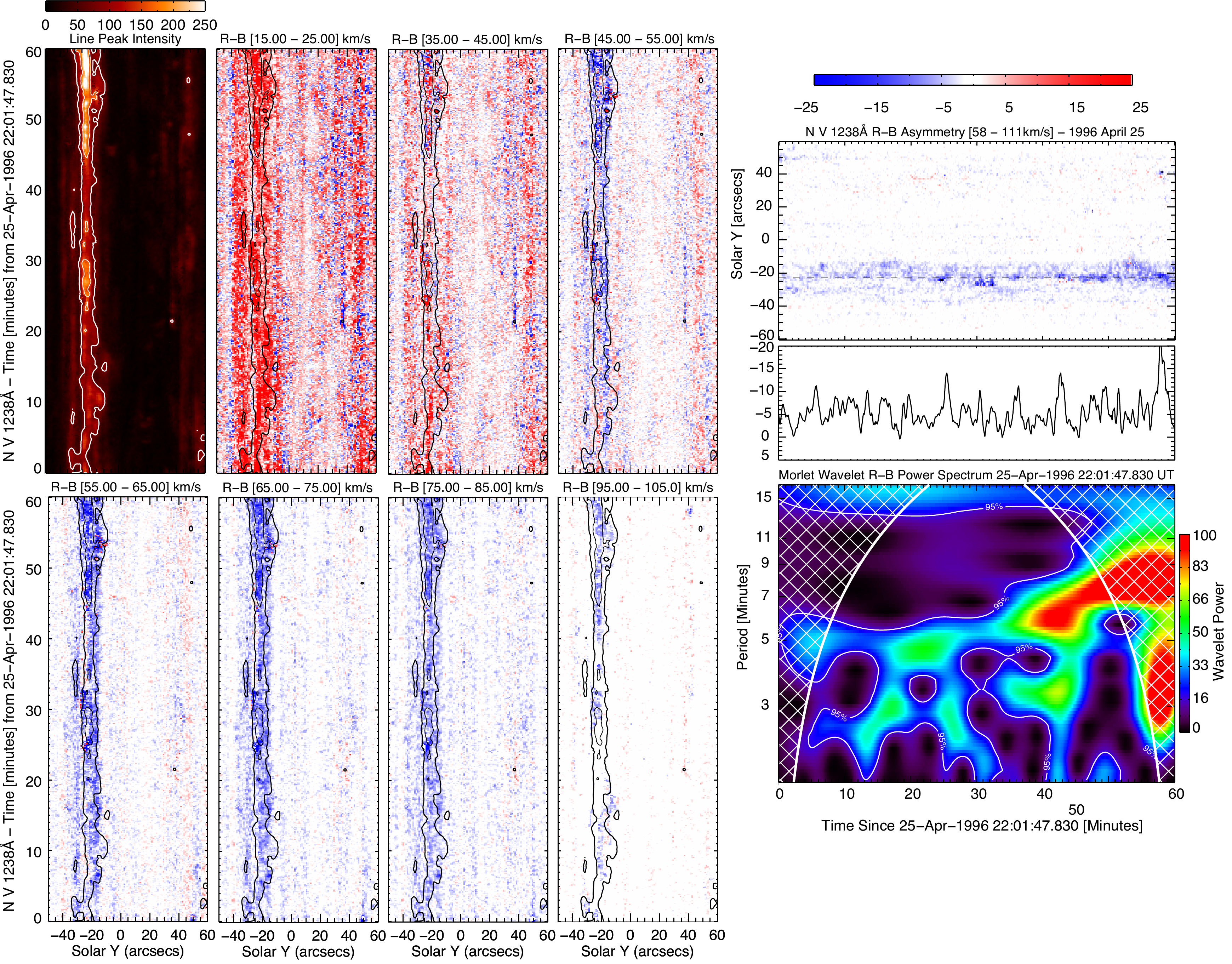}
	\caption{\label{fig:asymobssum} Timeseries in \ion{N}{5}~1238\AA\ at disk center, 
	in a quiet Sun region, with SUMER (left panel). Data from 1996, April 25, 22:01-23:01 UT
	(7.5s exposure) SUMER quiet Sun, disk center, on the left panel rows. The panels show the 
	peak line intensity and the R-B profile asymmetry in steps from 15 to 105~km~s$^{-1}$. 
	The panels are overlaid with intensity contours at peak intensity levels of 75 and 
	150 counts to delineate the supergranular network boundary (thick solid lines) 
	and the strongest network locations (thin solid lines) respectively. The right 
	column of panels shows the wavelet analysis of the \ion{N}{5} 1238\AA{} 
	RB (58-111~km~s$^{-1}$) timeseries at $y = -22$\arcsec .
	The bottom panel shows the cross-hatched cone of influence and 
	contours of the 90\% significance level in the timeseries.}
\end{figure*}

Figure~\ref{fig:asymobssum} shows SUMER observations of \ion{N}{5}~1238\AA\ 
of a quiet Sun and network region. The intensity of the line is slowly changing 
in the bright network region over the course of the one hour long timeseries. 
The eight RB panels show (once again), in the 
magnetized network, redward asymmetry for velocities less 
than 50~km~s$^{-1}$ shifting rapidly to blueward asymmetries for velocities in 
excess of 50~km~s$^{-1}$. In the internetwork regions very little 
asymmetry is found for these velocities. 

Further analysis of the \ion{N}{5}~1238 \AA{} time series shows that the 
characteristic timescale on which these upflow events seem to recur is of 
order 3-15 minutes. This is illustrated on the right side of 
Figure~\ref{fig:asymobssum} which shows, as a function of time and space, 
the episodic nature of the RB asymmetry of the \ion{N}{5} line (summed from 58 to 
111~km~s$^{-1}$). This is confirmed by a wavelet analysis 
(using the Morlet wavelet) of the RB timeseries: the resulting wavelet power 
spectrum (bottom right panel) shows significant (at 95\% significance level) 
power throughout the timeseries at periods of order 3-5 minutes, as well as 
some at 15 min. From the models we can not make this type of study 
since we were not able to simulate the recurrence of these events. The middle 
panel shows the RB timeseries at $y=-22$\arcsec{} with clear peaks of 
blueward asymmetry occurring, and typical events lasting of order 1-3 minutes. 
It is also interesting to see that the strong intensity lasts for more than 5 minutes 
for the event at $y=-22$\arcsec{}, i.e., a few minutes more than the RB asymmetry 
signal. This lifetime of the strong intensity signal seems to agree with the 
simulations and the SDO/AIA results as mentioned above. 

\begin{figure*}
	\includegraphics[width=0.99\textwidth]{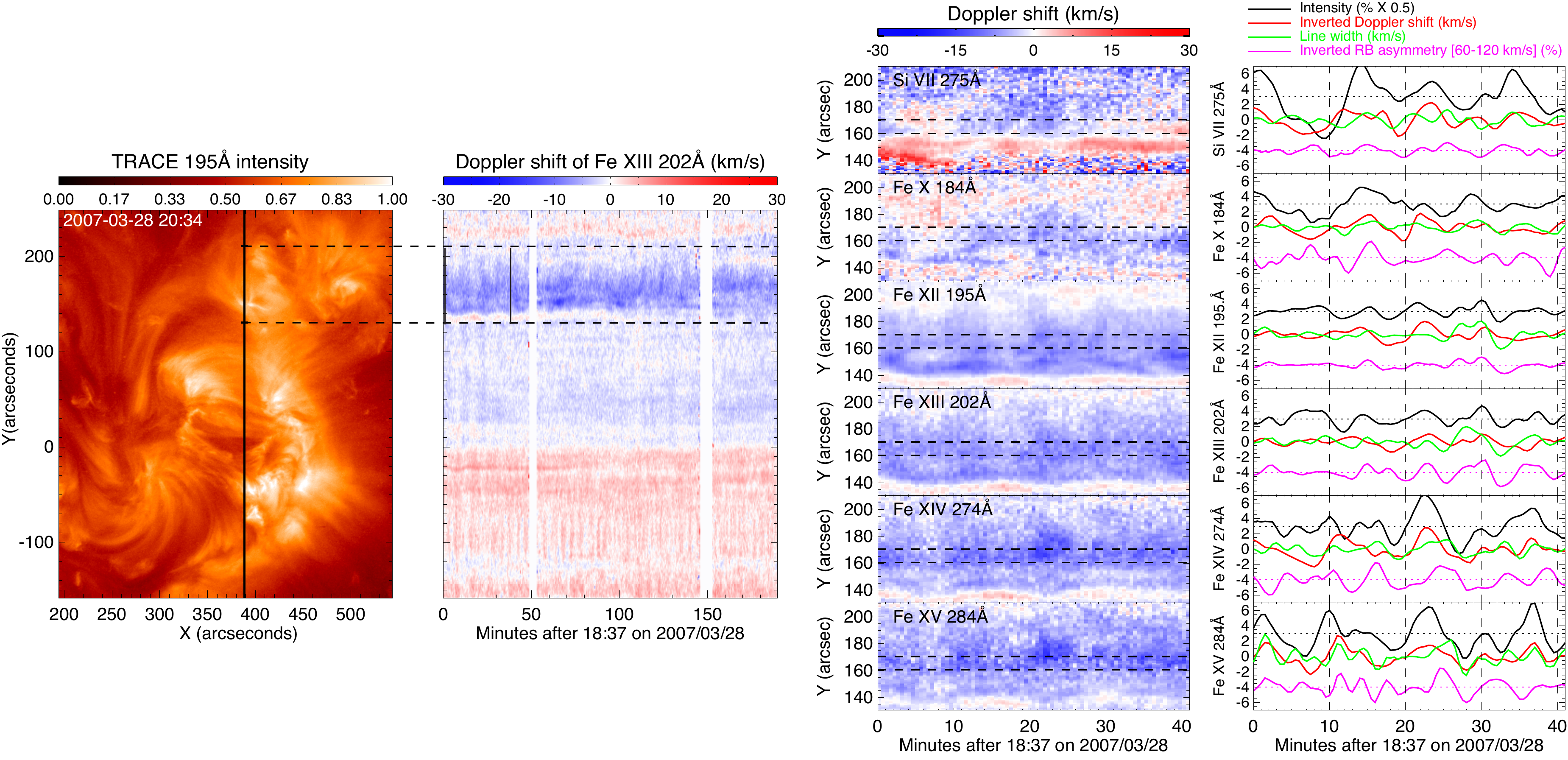}
	\caption{\label{fig:asymobseis} This figure is adapted from \citet{Tian:2012uq}: ``Temporal evolution of the Doppler shift of several emission lines in the range of $y \sim [130\arcsec ,210\arcsec]$ in left panel. Right panel: The black, red, green, and violet curves represent respectively the line intensity, Doppler shift (inverted), line width, and RB asymmetry (inverted) averaged over the region between the two dashed lines shown in the set of panels 	in the third column. The intensity has been normalized to the local background and is shown as the percentage divided by 2. The RB asymmetry has been normalized to the peak intensity of the line profile and is shown in the unit of percentage. The units of Doppler shift and line width are km~s$^{-1}$. For the purpose of illustration, the intensity and RB asymmetry are offset by 3 and -4 respectively on the $y$-axis". }
\end{figure*}

Figure~\ref{fig:asymobseis} shows EIS observations of several coronal lines of an 
active region. From this data set \citet{Tian:2012uq}
studied the recurrence of these signals. Here we are interested in the duration 
of a particular event. Around $t=20$min there is an event in the hotter lines that 
lasts for 5 minutes in intensity and Doppler shift, and the RB asymmetry and line 
width starts later and seems to last shorter (3-4 min).  Similarly around $t=15$min
for the cooler lines (\ion{Si}{7}~275\AA , \ion{Fe}{10}/XII~184\AA ), the intensity 
Doppler shift and RB asymmetries last for more than 3 minutes. Note the large 
difference of the lifetime of these events in the coronal lines ($\sim 5$min) 
and the transition region lines ($1-3$min). 

The lifetime of the RB asymmetries in the transition 
region lines as shown with SUMER in Figure~\ref{fig:asymobssum} is 
shorter (1-3 minutes) than the lifetime of the RB asymmetries and Doppler shifts 
of the coronal lines (Figure~\ref{fig:asymobseis}). This is very similar to what 
we see in the simulations. From the model, the transition region lines show large 
RB asymmetries for a lifetime of roughly 2-3 minutes, but the coronal lines 
shows large RB asymmetries for $\sim 5$ minutes as described Section~\ref{sec:rbdisk}.
Unfortunately we do not have access to RB asymmetry maps of an identical region 
from the observations for both temperature regimes (transition region and coronal lines). 
In addition we do not have simultaneous \Halpha\ observations to link the events with RBEs. 
Therefore further observational studies are needed.

\section{Discussion and conclusions}\label{sec:conclusions}

We have studied the thermodynamics of a plausible candidate for a type~II spicule 
model using chromospheric, transition region, and coronal synthetic observables 
and compared them with observations. The driver of this candidate spicule has been 
studied by \citet{Martinez-Sykora:2011uq} who found that it evolved naturally as a 
consequence of the magnetic field dynamics of the model (which include flux emergence). 
In short, chromospheric material is ejected into the corona as plasma is squeezed horizontally 
by a strong Lorentz force. The pressure increases and thus deflects the plasma and 
forces it to flow vertically along the magnetic field, most rapidly towards the low 
densities of the corona. Several synthetic observables reproduce the observations. 
However, mostly in the chromosphere, some synthetic properties differ from the 
observations. 

The chromospheric diagnostics seem harder to reconcile with the observations than 
coronal diagnostics for obvious reasons: the chromosphere is a transition between 
the photosphere and corona where a mixture of various physical processes play a role
such as, non-LTE, non-equilibrium ionization, ion-neutral interaction effects, etc. In addition, 
calculating the synthetic observables is more complex for chromospheric lines than 
for coronal lines. 
In addition to this complexity, it is also
clear that the simulated spicule is not very violent and shows
chromospheric densities that seem to be too low by at least an order
of magnitude. 
All of these issues likely help explain why the simulated spicule does not 
show a clear RBE signal: because of the low hydrogen density, the
opacity is 
too low above $z=2$~Mm so that not clear absorbing component in the wings of the 
synthetic profiles can be detected. Such a component is observed in RBEs on the Sun.
Even though we find large differences between the observed and synthetic profiles, 
specially as seen on the disk, it is important to note that one can deduce what is missing
 and/or could help to mimic the observables by studying the differences in detail. 
The low density indicates that not enough chromospheric material is ejected 
into the corona and/or, as a result of the simplified magnetic field 
configuration, the spicule expands too fast 
in the corona due to the field expansion. We note that despite
discrepancies between synthetic and real chromospheric observables, the physical 
properties such as the plasma being heated and accelerated along the spicule 
does seem to correspond with what is observed on the Sun.

The off-limb observables seem to show better agreement than the on-disk 
observables. The  
synthetic \cair\ and \Halpha\ emissions show structures and evolution
that are (in part) similar 
to the observations. Even though apparent upward velocities seem too
low, the chromospheric signal of the synthetic spicule fades in a few 
seconds (similar to observations). In the simulations this occurs as a result of being heated mainly by Joule heating, and to a lesser 
extent by thermal conduction and advection.  
However, the synthetic \cair\ and \Halpha\ emission are fainter than the observed 
ones, perhaps because the spicule does not eject enough chromospheric material.
It is interesting that we ``observe'' torsional behavior in the spicule that is similar to the 
observations \citep{De-Pontieu:2012bh}. This rotation is seen not only 
in chromospheric lines but also in transition region lines and to a lesser extent 
in the coronal lines. This rotation is generated in the chromosphere as a result of 
the compression which leads to a high magneto baroclinic term. 
It is interesting to note that all the chromospheric dynamics
of the spicule are at the low end of the observed range, i.e., both upflows and
rotation velocities are lower compared to observations by the same relative factor. 

The spicule shows emission in several transition region and coronal EUV lines 
(\hedi, \ion{Fe}{9}-XIV, etc). 
The transition region lines show emission at the limb 
following a path that is similar to a parabolic profile and coronal lines show a wave 
like propagation into the corona ($\sim 100$~km~s$^{-1}$). The former type of 
evolution is a result of various heating contributions, and it does not show downflows. 
The latter type of evolution is a combination of waves, flows and a thermal conduction 
front. These evolutions seem in agreement with the observations 
\citep{De-Pontieu:2011lr}. On the disk, two different brightenings have been 
observed in transition and coronal lines, one with the ejection of the spicule, and
another one that increases slowly with time over a range of a few hundred 
seconds. Some observational examples seems to have similar properties, but 
the brightenings from the observations are weaker than the synthetic ones. 
This difference may be a result of the magnetic field configuration, 
the computational domain is small and the heating on the Sun can be 
expanded to larger distances along the magnetic field lines than in our simulation. 

The transition region and coronal lines show not only blue Doppler shifts but also 
strong asymmetry and an increase of the line width in the spicule. In 
addition, we find a different time evolution of the properties of the profiles
for transition region lines and coronal lines. The increase in Doppler shift, 
line width and asymmetry lasts significantly longer in coronal lines. 
We show some observational examples of the duration 
of the Doppler shifts and RB asymmetries of transition region and coronal lines 
where the synthetic results seem to agree with the observations. However further
studies are needed because we were not able to compare transition region lines
with coronal lines in the same region and time on the Sun. In addition, in  the 
observations we can not link the RB asymmetries events  with RBEs because we 
did not have simultaneous \Halpha\ observations. We also want to point out that 
it is crucial to combine 
the information of the Doppler shift and the asymmetry of the line in order to
correctly determine the plasma velocities. Using only Doppler shift or only 
asymmetries is likely to lead to incorrect conclusions about plasma velocities. 
We also take into account the effects of the spectral broadening and spatial 
resolution into these results. Some of these lines may be impacted by the non-equilibrium 
ionization \citep{Judge:2012uq}.

Several processes still need to be taken into account that will likely impact the synthetic 
diagnostics. At a minimum we need to include ion-neutral coupling effects, and 
non-equilibrium ionization. The latter plays a role for the energy properties of the 
chromospheric plasma \citep{Leenaarts:2012cr}, but also in some transition 
region EUV line diagnostics \citep{kosovare:2012}. It is also crucial to increase
the spatial resolution of the simulations and keep in mind that the Joule heating in 
the model is dependent on numerical diffusivities and dissipation and that heating 
on the Sun behaves differently.

Many properties of type~II spicules, both from the observations and the model 
presented here, indicate that it cannot be modeled properly in 1D models. 
For instance, from the observational point of view, the spicules experience torsional 
motions. Many spicules seem to show multi-thread structuring. In the modeled 
type~II spicule the dominant acceleration and heating mechanisms vary significantly 
across the spicule, i.e., perpendicular to the magnetic field. For instance, the 
compression caused by the Lorentz force and the dissipated magnetic energy does 
not occur uniformly across the spicule. In fact, these processes vary in location and 
time, i.e., different "field lines" experience different mechanisms, and the dominant 
mechanism(s) at one location also evolve with time. The ejection and the expansion 
of the plasma is not uniformly distributed across the coronal part of the spicule. 
In addition, these "field lines" also experience changes in the magnetic connectivity, 
so both heating and acceleration jump from line to line in space and time. Most 
importantly, the spicule is not a structure following a confined magnetic flux tube. 
In short, all of these processes do not occur only along the field lines but also 
perpendicular to the field lines, something that is ignored in a 1D approach.

The chromosphere in our numerical simulation is significantly different from the 
solar chromosphere in several aspects: it is not as dynamic as revealed by 
observations, and type~II spicules occur much more frequently on the Sun. In fact, 
we only have a single event that resembles (in several aspects) type~II spicules. It 
is thus crucial to approach that deficiency in the models, and it needs to be investigated 
whether ion-neutral interactions, a more realistic treatment of the small-scale flux 
emergence, large scale motions or better spatial resolution can produce a more 
dynamic chromosphere. As mentioned above, the modeled ambient and emerging 
magnetic field is highly simplified. This may lead to the lack of dynamics and of 
type~II spicules. It is also important to mention that other physical processes, such 
as reconnection, could drive type~II spicules that might produce similar observables, 
but this needs to be investigated. Finally, type~II spicules are not occurring in the 
simulations as frequently as in the observations and the episodic nature of the RB 
asymmetry is missing in the models. A satisfactory model of spicules must be able 
to reproduce these observables too.

\section{Acknowledgments}

The research leading to these results has received funding from the 
European Research Council under the European Union's Seventh Framework 
Programme (FP7/2007-2013)/ERC Grant agreement no. 291058. 
We gratefully acknowledge support by NASA grants NNX08AH45G, 
NNX08BA99G, NNX11AN98G, NNM07AA01C (Hinode), and NNG09FA40C 
(IRIS). TMDP's research was supported by the NASA Postdoctoral Program 
at Ames Research Center through contract number NNH06CC03B. 
H. Tian's work at CfA is supported under contract 8100002705 from 
Lockheed-Martin to SAO. The 3D simulation and synthesis 
have been run on clusters from the Notur project, and the Pleiades cluster 
through the computing project s1061 from the High End Computing (HEC) division of NASA. 
We thankfully acknowledge the computer and supercomputer 
resources of the Research Council of Norway through grant 170935/V30 
and through grants of computing time from the Programme for 
Supercomputing. The Swedish 1-m Solar Telescope is operated by the Institute for Solar Physics of the Royal Swedish Academy of Sciences in the Spanish Observatorio del Roque de los Muchachos of the Instituto de Astrof\'{\i}sica de Canarias.  

To analyze the data we have used IDL and Vapor 
(http://www.vapor.ucar.edu). 

Eamon Scullion acquired the SST observations.

\bibliographystyle{aa}

\end{document}